\documentclass[twocolumn,prl,aps,superscriptaddress, longbibliography]{revtex4-1}
\usepackage[T1]{fontenc}
\usepackage[latin9]{inputenc}

\usepackage{amsmath}
\usepackage{amssymb}
\usepackage[usenames,dvipsnames]{xcolor}
\usepackage{bbm}
\usepackage{braket}
\usepackage{mathrsfs}  
\usepackage{mathtools}
\allowdisplaybreaks

\usepackage{comment}

\usepackage{graphicx}

\usepackage{textcmds}

\usepackage[colorlinks=true]{hyperref}  
\hypersetup{
    bookmarks=true,         
    unicode=false,          
    pdftoolbar=true,        
    pdfmenubar=true,        
    pdffitwindow=false,     
    pdfstartview={FitH},    
    pdftitle={Composite fermion duality},    
    pdfauthor={Ruegg, Chaudhary, Slager},     
    pdfsubject={},   
    pdfcreator={},   
    pdfproducer={}, 
    pdfkeywords={} {} {}, 
    pdfnewwindow=true,      
    colorlinks=true,       
    linkcolor=blue, 
    citecolor=blue,        
    filecolor=magenta,      
    urlcolor=blue,           
	breaklinks=true
}

\definecolor{orange}{rgb}{1,0.5,0}

\newcommand{\sect}[1]{\vspace{0.3em}{\it #1.}---}

\usepackage{bbold}

\newcommand{\be}{\begin{equation}}
\newcommand{\ee}{\end{equation}}
\newcommand{\bea}{\begin{eqnarray}}
\newcommand{\eea}{\end{eqnarray}}

\newcommand{\sign}{\,\text{sign}}

\renewcommand{\vec}[1]{\boldsymbol{#1}}

\newcommand{\cd}{c^{\dagger}}

\newcommand{\psiup}{\psi_{\uparrow}}
\newcommand{\psidown}{\psi_{\downarrow}}
\newcommand{\psid}{\psi^{\dagger}}

\newcommand{\zup}{z^{\uparrow}}
\newcommand{\zdown}{z^{\downarrow}}

\newcommand{\wdown}{w^{\downarrow}}

\newcommand{\psit}{\psi_{\text{t}}}
\newcommand{\psib}{\psi_{\text{b}}}
\newcommand{\psitd}{\psi_{\text{t}}^{\dagger}}
\newcommand{\psibd}{\psi_{\text{b}}^{\dagger}}

\newcommand{\psitupd}{\psi_{\text{t}\uparrow}^{\dagger}}
\newcommand{\psibupd}{\psi_{\text{b}\uparrow}^{\dagger}}

\newcommand{\psitdownd}{\psi_{\text{t}\downarrow}^{\dagger}}
\newcommand{\psibdownd}{\psi_{\text{b}\downarrow}^{\dagger}}

\newcommand{\fsl}[1]{\ensuremath{\mathrlap{\!\not{\phantom{#1}}}#1}}

\begin{document}

\title{On dualities of paired quantum Hall bilayer states at $\nu_T = \frac{1}{2} + \frac{1}{2}$}
\author{Luca R\"uegg}
\email{lr537@cam.ac.uk}
\address{TCM Group, Cavendish Laboratory, University of Cambridge, J. J. Thomson Avenue, Cambridge CB3 0HE, United Kingdom}
\author{Gaurav Chaudhary}
\email{gc674@cam.ac.uk}
\address{TCM Group, Cavendish Laboratory, University of Cambridge, J. J. Thomson Avenue, Cambridge CB3 0HE, United Kingdom}
\author{Robert-Jan Slager}
\email{rjs269@cam.ac.uk}
\address{TCM Group, Cavendish Laboratory, University of Cambridge, J. J. Thomson Avenue, Cambridge CB3 0HE, United Kingdom}

\date{\today}
\begin{abstract} 
Density-balanced, widely separated quantum Hall bilayers at $\nu_T = 1$ can be described as two copies of composite Fermi liquids (CFLs). 
The two CFLs have interlayer weak-coupling BCS instabilities mediated by gauge fluctuations, the resulting pairing symmetry of which depends on the CFL hypothesis used. 
If both layers are described by the conventional Halperin-Lee-Read (HLR) theory-based composite electron liquid (CEL), the dominant pairing instability is in the $p+ip$ channel; whereas if one layer is described by CEL and the other by a composite hole liquid (CHL, in the sense of anti-HLR), the dominant pairing instability occurs in the $s$-wave channel. 
Using the Dirac composite fermion (CF) picture, we show that these two pairing channels can be mapped onto each other by particle-hole (PH) transformation.
Furthermore, we derive the CHL theory as the non-relativistic limit of the PH-transformed massive Dirac CF theory.
Finally, we prove that an effective topological field theory for the paired CEL-CHL in the weak-coupling limit is equivalent to the exciton condensate phase in the strong-coupling limit.
\end{abstract}

\maketitle
\sect{Introduction}
Dualities constitute a powerful approach in theoretical physics. 
Ranging from the Kramers-Wannier duality~\cite{Kramers1941,Kogut1979} to the celebrated AdS/CFT correspondence~\cite{Maldacena_1999,Witten1998}, they offer a versatile tool to analyze the full content of a variety of theories and systems~\cite{beekman2017}, including disordered spin systems and strongly interacting conformal field theories.
As dualities in essence are non-perturbative, they find particular applications in strongly interacting systems, often mapping strongly coupled field theories to their dual weakly coupled counterparts~\cite{Seiberg2016}. 
As such, they provide a deeper insight into the novel interplay of topology and strong correlations in fractional quantum Hall physics~\cite{Seiberg2016,Wang2015,Sodemann2017}.

Quantum Hall bilayers (QHBs) at filling $\nu_T = \nu_{\uparrow} + \nu_{\downarrow} = \frac{1}{2} + \frac{1}{2}$ ($\uparrow/\downarrow$ is the upper/lower layer index) are an intriguing example of strongly correlated physics. 
At zero layer separation $d$, interlayer Coulomb interactions ensure that every electron from one layer is aligned with a hole from the other layer. 
These macroscopically large numbers of excitons occupy the unique Halperin~(111) state~\cite{Halperin1983} to form an exciton condensate (XC) phase~\cite{Spielman2000,Eisenstein2004}. 
At $d \gg \ell$, where $\ell$ is the magnetic length $\sqrt{\hbar / (eB)}$, the two layers decouple and are believed to form two isolated $\nu = \frac{1}{2}$ composite Fermi liquids (CFLs)~\cite{Halperin1993}. 
The phase of the QHB between these two limits is a long-debated topic with several possibilities discussed in the literature~\cite{Bonesteel1993,Bonesteel1996,Morinari1999,Kim2001,Simon2003,Shibata2006,Doretto2006,Zhu2017,Lian2018}. 
It was realized quite early that an effective interlayer attraction exists for composite fermions (CFs) at all layer separations~\cite{Bonesteel1993, Bonesteel1996}. 
This was later extended to analytically~\cite{Cipri2014,Isobe2017} and numerically~\cite{Moller2008,Moller2009} show that the dominant BCS pairing occurs in the $l = 1$ angular momentum channel, leading to a $p+ip$ pairing instability~\footnote{In the Supplemental Material, we show that the dominant channel rather corresponds to $l = -1$. This is, however, only a semantical distinction and we agree with the results of the aforementioned references.}.

Alternatively, it was recently hypothesized that in the weak-coupling limit, the bilayers form $s$-wave pairs of a composite electron liquid (CEL) in one layer and a composite hole liquid (CHL) in the other layer~\cite{Liu2022}. 
This point of view is particularly appealing to understand the experimental signatures of preformed excitons before the onset of the XC phase~\cite{Liu2022,Eisenstein2019}, since it creates intuitively correct quasiparticles in the weak-coupling limit that can be linked to the strong-coupling XC phase via a BCS-BEC crossover. 
An excellent numerical agreement with a trial wave function for an $s$-wave CEL-CHL (CEL-CHL-s) paired state was also found~\cite{Wagner2021}. 
Using Eliashberg theory, it was analytically shown that the $s$-wave is the dominant pairing channel in the CEL-CHL description~\cite{Rueegg2023}, which is also in agreement with the $s$-wave paired dipole quasiparticles description~\cite{Predin2023, Predin2023a}. 
Indeed, the BCS-BEC crossover to the XC phase was argued in Ref.~\cite{Sodemann2017}, albeit using the $p+ip$ paired CEL-CEL (CEL-CEL-p) in the weak-coupling limit. 

Interestingly, the two weak-coupling scenarios, the CEL-CHL-s state and the CEL-CEL-p state, possess nearly identical overlap with the exact diagonalization (ED) ground state~\cite{Wagner2021}. 
Furthermore, at zero temperature, these two paired states have the same pairing strength~\cite{Rueegg2023}. 
A recent modification of the Eliashberg theory further showed that the CEL-CHL-s and CEL-CEL-p pairing strength can be made close to degenerate, or under the inclusion of a further term exactly degenerate~\cite{Lotric2024}. 
These observations suggest a close relation between the CEL-CEL-p and CEL-CHL-s paired states. 
As noted before, a topological equivalence between the weak-coupling CEL-CEL-p state and the strong-coupling Halperin~(111) state has already been argued before~\cite{Sodemann2017}. 
Similarly, is the strong-coupling XC phase dual to the weak-coupling CEL-CHL-s state? 
Moreover, how are the two weak-coupling descriptions CEL-CHL-s and CEL-CEL-p related to each other, given their different pairing structures? 
In this work, we address these questions.

\begin{figure}
\includegraphics[width=0.45\textwidth]{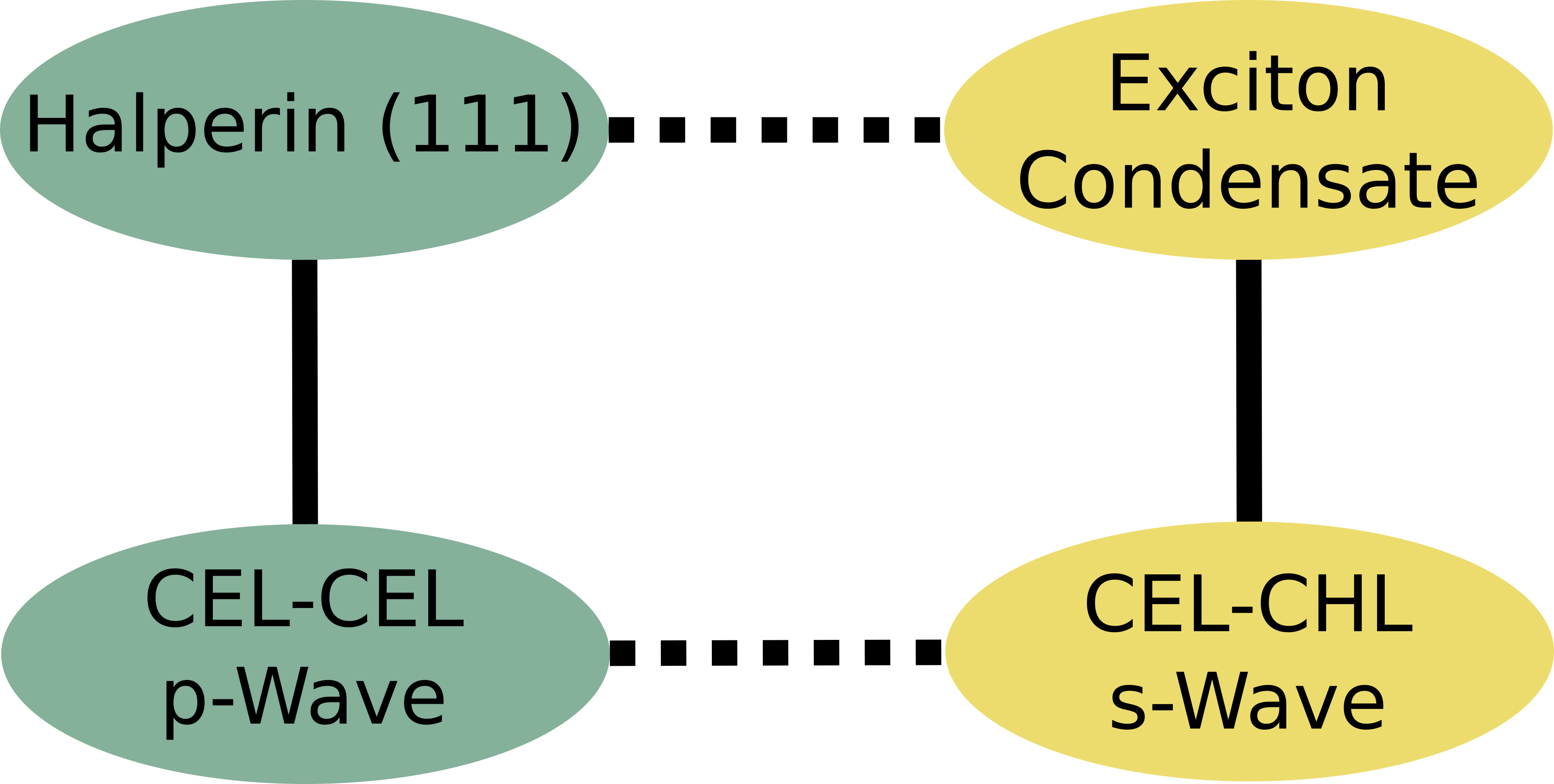}
\caption{Diagram of the main dualities. Solid lines stand for topological equivalence, and dashed lines are for particle-hole duality.}
\label{fig: dualities summary}
\end{figure}

Here, we establish a weak-coupling correspondence between the two pictures based on field-theoretical arguments. 
We start with the Dirac CF theory~\cite{Son2015} which contrary to the Halperin-Lee-Read (HLR) theory is manifestly particle-hole (PH) symmetric~\cite{Halperin1993}.  
Nonetheless, in some cases the low-energy predictions of the HLR theory have been shown to have an emergent PH symmetry~\cite{Wang2017,Cheung2017}, indicating a deeper connection between these two theories. 
We consider a bilayer Dirac CF theory with $s$-wave layer-triplet spinor-singlet pairing.
By introducing a mass in the Dirac particles and taking their non-relativistic limit, the CF Fermi surface projected paired state is mapped to a CEL-CEL-p state with layer-triplet pairing. 
We show that in the bilayer Dirac CF theory when one of the layers is PH transformed, the pairing term becomes singlet in the layer pseudospin and triplet in the spinor, while keeping its $s$-wave structure. 
Taking the non-relativistic massive limit and projecting over the CF Fermi surface, we arrive at the CEL-CHL-s state with interlayer-singlet pairing.
Finally, for the CEL-CHL-s state, we write down an effective topological field theory which is equivalent to the XC theory described by the Halperin~(111) $K$-matrix~\cite{Wen1992,Wen1993,Sodemann2017}. 
A summary of these dualities is presented in Fig.~\ref{fig: dualities summary}.
We further present wave function arguments that circumstantiate these results.

\sect{Weak-coupling duality} 
Eliashberg calculations under a random-phase approximation in the CEL-CEL theory show that the leading order dependency of the pairing strength on the angular momentum channel $l$ goes like $l^2 -2l$~\cite{Isobe2017}.
Hence, the $l+1$ and $-l+1$ channels are degenerate, and the $l=1$ channel has the strongest pairing.
The same calculations in the CEL-CHL theory show that there the leading order of the pairing strength goes like $l^2$~\cite{Rueegg2023}, thus exhibiting an effective time-reversal symmetry (TRS).
As a result, $\pm l$ pairing channels are degenerate, and $l=0$ leads to the most dominant pairing instability.
At zero temperature (Matsubara frequency $\omega_m = 0$), the CEL-CEL $l+1/-l+1$ and the CEL-CHL $\pm l$ channels all have the same pairing strength.
This indicates a PH-duality for interlayer-paired ground states, where the pairing symmetry increases by one when going from the CEL-CHL to the CEL-CEL description.

To establish a duality between the CEL-CEL-p and CEL-CHL-s weak-coupling scenarios, we take the relativistic Dirac CF theory as our starting point. 
The non-relativistic HLR theory~\cite{Halperin1993} (denoted as CEL here) and anti-HLR theory (denoted as CHL)~\cite{Barkeshli2015} are not manifestly PH symmetric~\cite{Girvin1984,Kivelson1997,Rezayi2000,Barkeshli2015,Balram2016,Geraedts2016}.
At the mean-field level at $\nu = \frac{1}{2}$, non-relativistic CFs are charge-neutral and have a quadratic dispersion; obscuring the notion of PH symmetry. 
The relativistic Dirac CF theory is manifestly PH symmetric~\cite{Son2015}. 
The microscopic starting point for the Dirac CF theory are relativistic Landau levels at half-filling (charge neutrality).
The Dirac CF theory is then derived by using the 2+1D fermionic particle-vortex duality~\cite{Seiberg2016}. 
The CF nature of the emergent quasiparticles can be seen from the fact that under the particle-vortex transformation particle density and magnetic field are interchanged.
Similar to the HLR theory, the Dirac CF theory describes a system with non-zero density but a vanishing magnetic field. 
Indeed, the HLR Lagrangian can be obtained from the Dirac CF Lagrangian by introducing a PH-symmetry-breaking mass term and integrating out the valence band of the massive Dirac spectrum in the limit of large chemical potential in the conduction band~\cite{Son2015,supplementary}.

As discussed in Ref.~\cite{Sodemann2017}, consider a bilayer of Dirac CFs with an interlayer BCS pairing of the type $i \Delta \psi \sigma_y \tau_x \psi$, where $\psi = (\psit, \psib)$ denotes the Dirac CF fields in the top/bottom layer, and $\tau_i / \sigma_i$ are the Pauli matrices acting on the layer/spinor space.
The gap function $\Delta$ is homogeneous with $s$-wave symmetry. 
The choice of singlet pairing structure in the internal pseudospin of the Dirac field requires a triplet structure in the layer pseudospin to ensure full fermionic anti-symmetry of the CF Cooper pairs.  
The explicit Bogoliubov-de Gennes (BdG) Hamiltonian is block-diagonal with two identical blocks that are
\begin{equation}
    \mathcal{H}_{\text{BdG}}^{\text{tb}} =
    \begin{pmatrix}
        \vec k \cdot \vec\sigma- \mu & -i\Delta \sigma_y\\
        i\Delta \sigma_y & \vec k \cdot \vec\sigma^{\ast} + \mu
    \end{pmatrix}
\end{equation}
in the basis $(\psi_{\text{t/b}}(\vec k), \psi^{\dagger}_{\text{b/t}}(-\vec k))^{\intercal}$, and $\mu$ is the chemical potential. 
Here, we assume that each layer is exactly at the LL filling factor $\nu = \frac{1}{2}$ and for now ignore the minimal coupling of the momentum $\vec{k}$ to the statistical gauge fields $\vec{a}_{\text{t/b}}$. 
Diagonalizing the single-particle part and then projecting onto the Fermi surface, assuming $\mu \gg |\Delta|$, the pairing term becomes $\Delta e^{i\theta_{\vec k}} \, \psid_{\text{t}+}(\vec k) \psid_{\text{b}+}(-\vec k)$, where $\theta_{\vec k}$ is the polar angle of $\vec k$, and $+$ denotes the upper Dirac eigenstate~\cite{supplementary}.
Therefore, the pairing projected over the Fermi surface acquires a $p_x + ip_y$ form with a triplet structure in the layer pseudospin. 
This is identical to the celebrated Fu-Kane model for chiral topological superconductivity on the surface of a topological insulator~\cite{Fu2008}. 
In the present case, the only differences are: (i) the superconductor describes a pair of CFs, and (ii) it has a double degeneracy due to the presence of an identical BdG copy under reversing the layer degree. 
Now, we can take the non-relativistic limit by introducing a Dirac mass in the CFs and integrating out the lower single-particle band~\cite{Son2015,supplementary}. 
The underlying CF theory becomes the conventional HLR theory of the CEL, while the pairing structure remains invariant. 
As a result, the CEL-CEL-p state is recovered starting from the $s$-wave paired Dirac CFs.

Now, from the same starting point of paired Dirac CFs, we obtain the CEL-CHL-s state. For this purpose, we PH transform the bottom layer which has the following effect on the pairing term~\cite{supplementary}
\begin{equation}
    i \Delta \psi \sigma_y \tau_x \psi \quad \rightarrow \quad i  \Delta \psi \sigma_0 \tau_y \psi.
\end{equation}
The interlayer BdG Hamiltonian is
\begin{equation}
    \mathcal{H}_{\text{BdG}}^{\text{tb}} =
    \begin{pmatrix}
        \vec k \cdot \vec\sigma- \mu & -\Delta \sigma_0\\
        -\Delta \sigma_0 & \vec k \cdot \vec\sigma^{\ast} + \mu
    \end{pmatrix}.
\end{equation} 
At the Fermi surface the pairing term becomes $-\Delta \, \psid_{\text{t}+}(\vec k) \psid_{\text{b}-}(-\vec k)$~\cite{supplementary}. 
We now take the non-relativistic limit by introducing a Dirac mass, while keeping in mind that the PH-transformation in the bottom layer switches the sign of the mass term $\bar\psib \psib$. 
For the non-relativistic limit, we now have $\psi_{\text{b}-}$ for the Dirac fermions above the Fermi surface because the PH transformation inverts the Dirac cone in the bottom layer.
Thus, we obtain the desired $s$-wave pairing along with the underlying CF theory described by the CEL in the top layer and the CHL in the bottom layer~\cite{supplementary}.
This mechanism can also be understood from the Berry phase argument.
For the Berry curvature $\mathcal{F}_{\text{b}}(\vec k)$ in the bottom layer we have: $\mathcal{CT} \, \mathcal{F}_{\text{b}}(\vec k) \, (\mathcal{CT})^{-1} = - \mathcal{F}_{\text{b}}(-\vec k)$.
Upon adding this to the Berry curvature of the top layer particle, the relative motion degree of freedom acquires no total Berry phase upon encircling the Dirac cone; contrary to the CEL-CEL case where the $\pi$ Berry phase can be seen to induce the $p$-wave pairing symmetry.

The weak-coupling duality of the interlayer BCS pairing carries over to higher angular-momentum symmetries.
For bilayer Dirac CFs with interlayer pairing of the form $\Delta e^{i l \theta_{\vec k}}$, where $l$ is the angular momentum channel, 
the appropriate pair wave function for even (odd) values of $l$ is $ \Delta e^{i l \theta_{\vec k}} \psi i \sigma_y \tau_{x(y)} \psi$.
A simple extension of the above BdG calculations shows that projection onto the Fermi surface for the CEL-CEL case acquires the form $\Delta e^{i (l+1) \theta_{\vec k}} \psi_+ \tau_{x(y)} \psi_+$ for even (odd) values of $l$.
PH transforming the bottom layer transforms the interlayer pairing as 
\begin{align}
     \Delta e^{i l \theta_{\vec k}} \psi i\sigma_y \tau_{x (y)} \psi \, &\rightarrow \,  (-1)^l \Delta e^{i l \theta_{\vec k}} \psi i \sigma_0 \tau_{y (x)} \psi ,
\end{align}
for even (odd) $l$~\cite{supplementary}.
This shows the duality between the CEL-CEL $l+1$ and CEL-CHL $l$ channels.

The above analysis shows a duality between CEL-CEL-p and CEL-CHL-s, which are two BCS weak-coupling pictures for the quantum Hall bilayers. 
However, it relies on obtaining the underlying CEL and CHL theories starting from the relativistic Dirac CF theory and introducing a PH-symmetry-breaking Dirac mass.  
Therefore, more precisely, our analysis shows that if the Dirac CF theory and the conventional HLR theory are in the same universality class, the resultant interlayer BCS states CEL-CEL-p and CEL-CHL-s are dual to each other. 
Recent developments have indicated that even though the HLR theory does not have an explicit PH symmetry, some of its response functions can be PH symmetric, thus opening a possibility for an emergent PH symmetry~\cite{Wang2017, Cheung2017}.  
Therefore, the Dirac CF and HLR theory may indeed be in the same universality class. However, a fully rigorous proof of this is still lacking.

\sect{Effective field theory} 
We now show a duality between the weak-coupling CEL-CHL-s and the strong-coupling XC phase. 
For this purpose, we derive an effective field theory in the weak-coupling limit and show its equivalence with the XC $K$-matrix theory in the BEC limit~\cite{Wen1992,Wen1993}.
Consider the condensate of Cooper pair bosons made out of the paired CEs and CHs given by the Lagrangian
\begin{align}
    \mathcal{L} = \frac{1}{2M} | iD_{\mu} \phi|^2 -\frac{\lambda}{4} (\phi^{\dagger} \phi)^2 -\frac{m^2}{2} \phi^{\dagger}\phi - a_+^{\mu} j_{\mu}^{\text{qp}} + \dots.
\end{align}
Here, $\phi$ is the CF Cooper pair field representing the CEL-CHL condensate,  $iD_{\mu} = i\partial_{\mu} - 2 a_{+\mu} $, $a_+  = \frac{a_{\text{t}} + a_{\text{b}}}{2}$, and $j^{\text{qp}}$ is the current of Bogoliubov quasiparticles.  
Because $a_+$ couples minimally to the condensate field, like an external field,  it creates vortex excitations in the condensate. 
The phase $\varphi$ of the condensate field $\phi$ winds by $2\pi$ around the vortex. 
Therefore, in the presence of vortices, we decompose $\varphi = \varphi_{\text{reg}} + \varphi_{\text{v}}$, where $\varphi_{\text{v}}$ has a singularity at the vortex core. 
Performing a singular gauge transformation $\vec{a}_+ \rightarrow \vec{a}_+ -\vec{\nabla} \varphi_{v}/2 $ and integrating out $|\phi|$, we obtain
\begin{align}\label{eq: abelian higgs lagrangian}
    & \mathcal{L} = - \beta_{\mu} (j^{\text{v}\mu} - \frac{1} {\pi} \epsilon^{\mu\nu\gamma} \partial_{\nu} \alpha_{\gamma} ) + \frac{2 m^2}{M\lambda} ( a_{+\mu} - \alpha_\mu  )^2 \notag\\
    &\hspace{2cm} -a_{+}^{\mu} j^{\text{qp}}_{\mu} +  \dots.
\end{align}
Here, the constraint of the conserved vortex current 
\begin{align}
    j^{\text{v}\mu} = \frac{1}{\pi} \epsilon^{\mu\nu\gamma} \partial_{\nu}\alpha_{\gamma},
\end{align}
where $\alpha_{\mu} = \partial_{\mu} \varphi_{\text{v}}$, is encoded in the Lagrange multiplier field $\beta$. Shifting $a_+ \rightarrow a_+ + \alpha$ and integrating out the massive $a_+$ field, an effective topological field theory takes the form
\begin{align}\label{eq: tqft bcs}
    \mathcal{L}_{\text{sc}} = \frac{1}{\pi} \beta d \alpha - \beta^{\mu} j^{\text{v}}_{\mu} - \alpha
    ^{\mu} j^{\text{qp}}_{\mu}.
\end{align}
An $s$-wave BCS condensate has gapped BdG and vortex excitations. 
The first term is the \qq{BF} topological field theory previously discussed for $s$-wave superconductors~\cite{Hansson2004}. 
The BF term encodes the $\pi$-phase accumulated when a Bogoliubov quasiparticle goes around the vortex or vice versa. 

In a gapped superconductor, the low-energy BdG excitations are an equal superposition of an electron and a hole, they are \textit{almost} charge neutral to the external field~\cite{Kivelson1990}. 
At higher energies, either the electron or the hole content starts to dominate and the quasiparticles acquire charge. 
While the Bogoliubov quasiparticle does not carry a charge current, if the superconductor is a spin-singlet, it does carry a spin current. 
Thus, this leads to a \qq{spin-charge} decoupling. 
In the BCS state described by the pairing of CEL and CHL, the order parameter vortex binds a half-quantized flux of the $a_+$ field.
Similar to the superconducting case, the composite BdG excitations are approximately neutral to $a_+$. 
However, the composite Bogoliubov quasiparticles have a unit charge with respect to the $a_- = \frac{a_{\text{t}} - a_{\text{b}}}{2}$ field. 
Based on this observation, we make the substitutions $\alpha \rightarrow a_-, \, \beta \rightarrow \frac{a_+}{2}$, so that the BF term becomes $\frac{1}{2\pi} a_+ da_-$.
The total topological field theory is the BF term for the $s$-wave superconductor plus the CS terms of the two CF fields and the external gauge field~\cite{Rueegg2023}
\begin{align}
\begin{split}
\label{eq: tqft}
    \mathcal{L}_{\text{top}} = \, &\frac{1}{2\pi} a_+ da_- - \frac{1}{8\pi} a_{\text{t}} d a_{\text{t}} + \frac{1}{8\pi} a_{\text{b}} d a_{\text{b}} + \frac{1}{4\pi} AdA.
\end{split}
\end{align}
The first three terms exactly cancel, which indicates the gapless neutral interlayer mode. 
The remaining CS term for the external field encodes the response of an interlayer $\nu = 1$ quantum Hall mode~\cite{Blok1990}.
Hence, we obtain the required effective topological field theory consistent with the $K$-matrix of the strong-coupling XC phase~\cite{Wen1992}.

\sect{Wave functions} As argued in Ref.~\cite{Sodemann2017}, an $l=1$ interlayer pairing wave function of CFs of the form
\begin{equation}
    \det \left[ \left( \frac{1}{(\zup_i - \zdown_j)^{\ast}} \right)_{ij} \right] \, \prod_{i < j}(\zup_i - \zup_j)^2 \prod_{i < j}(\zdown_i - \zdown_j)^2,
\end{equation}
is proportional (up to amplitude factors in the relative coordinates, which would also absorb the the interlayer singularities) to the Halperin~(111) wave function~\cite{Sodemann2017,supplementary}.
Here $z^{\uparrow / \downarrow} = x^{\uparrow / \downarrow} + iy^{\uparrow / \downarrow}$ are the coordinates of the particles in the upper/lower layer.
This can be taken as further evidence for the small-$d$ and large-$d$ limit to be connected since the location of zeros in the wave function does not change, only how they are approached. 

On the electron-hole side, the $d=0$ wave function already is an $s$-wave pairing function~\cite{Yang2001,supplementary}
\begin{equation}
\label{eq: xc}
    \psi_{\text{XC}} = \det \left[ \left( e^{\zup_i (\wdown_j)^{\ast} / 2\ell^2} \right)_{ij} \right].
\end{equation}
If we now just added the Jastrow factors of the CF construction, the resulting wave function describing the weak-coupling limit would not have the same structure of zeros as the strong-coupling $\psi_{\text{XC}}$.
To remedy this, we consider
\begin{align}
\label{eq: cel-chl s-wave}
\begin{split}
    &\psi_{\text{CEL-CHL,s}} = \prod_{i<j} (\zup_i - \zup_j)^2 \prod_{i<j} (\wdown_i - \wdown_j)^{\ast 2} \times \\
    & \det \left[ \left( \prod_{k \neq i}(\zup_i - \zup_k)^{\ast} \prod_{k \neq j}(\wdown_j - \wdown_k) \, e^{\zup_i (\wdown_j)^{\ast} / 2\ell^2} \right)_{ij} \right].
\end{split}
\end{align}
In this way of rewriting, the interlayer pairing symmetry of the expression inside the determinant is not affected by the presence of Jastrow factors. Therefore, because of the Jastrow factors outside the determinant, this wave function describes $s$-wave paired CEs and CHs.
Since, the Jastrow factors in the determinant cancel the ones in front, the phase structure is the same as for $\psi_{\text{XC}}$~\cite{supplementary}.

These arguments should be taken with a grain of caution though.
The wave functions are all implicitly assumed to be projected into the LLL, i.e. $z^{\ast} \rightarrow \partial_z, \, w \rightarrow \partial_{w^{\ast}}$, and all partial derivatives are acting from the left.
In general, LLL projection introduces additional zeros.
For example, $\psi_{\text{CEL-CHL,s}}$ and $\text{P}_{\text{LLL}} \psi_{\text{CEL-CHL,s}}$ do not have the same phase structure anymore~\cite{supplementary}.
Physically though, crossover from the CEL-CHL-s state to the XC state has a much simpler interpretation.  
As the strong-coupling limit is approached, the momentum-space pairing becomes a real-space pairing, resulting in $\zup_i \rightarrow \wdown_i$, and thus the phase of the Jastrow factors gets cancelled~\cite{Liu2022}.
This simple picture of the BCS-BEC crossover, however, is not possible when starting from the CEL-CEL-p state in the weak-coupling limit.

\sect{Discussion and conclusion} 
We have shown a weak-coupling duality between the CEL-CEL-p and CEL-CHL-s states; two possible descriptions for the BCS pairing state of CFs for widely separated QHBs at $\nu_T = \frac{1}{2} + \frac{1}{2}$.
For this duality, we derive the CEL-CEL-p theory from a bilayer Dirac CF theory with an $s$-wave pairing by projecting the BdG equations onto the Fermi surface and then taking the non-relativistic limit.
PH-transforming one layer in the bilayer Dirac CF theory, and applying the same BdG projection and non-relativistic limit procedure, we arrive at the CEL-CHL-s.
This weak-coupling duality generalizes to a PH-duality between CEL-CEL $l+1$ and CEL-CHL $l$ pairing, and coincides with observed numerical~\cite{Wagner2021} and analytical~\cite{Rueegg2023,Lotric2024} results.
Next, we derive an effective topological field theory for the CEL-CHL-s state and show its equivalence to the $K$-matrix theory of the strong-coupling XC phase in QHBs. 
This shows that the weak-coupling CEL-CHL-s state can adiabatically crossover to the strong-coupling XC phase.

We finally speculate how these dualities could help to understand different aspects of QHBs. 
The signatures of a BCS-BEC crossover to the XC phase were reported in recent experiments~\cite{Eisenstein2019, Liu2022}. 
In a full numerical many-body calculation as a function of separation $d$, such a crossover was seen in both the CEL-CEL-p and CEL-CHL-s as the starting weak-coupling state~\cite{Wagner2021}. 
However, many-body calculations are often difficult to directly connect with the experimental observables such as conductivity and tunneling spectroscopy. 
Therefore, a reliable mean-field theory is highly desirable for such detailed calculations. 
At the mean-field level, similar to the well-known BCS-BEC crossover, the CEL-CHL-s state can crossover to the XC as interlayer coupling is increased as $d$ is decreased. 
On the other hand, at this mean-field level, the CEL-CEL-p state will not go through a crossover but instead will encounter a phase transition. 
This can be readily seen by considering the BdG excitation spectrum of a $p$-wave pairing state. 
If we associate the BCS-BEC crossover with the change in the sign of the chemical potential of CFs that undergo pairing, as the coupling gets stronger and chemical potential reaches the band bottom, the $p+ip$ excitation spectrum closes the gap, where it encounters a phase transition before reaching the BEC regime. 
Therefore, to capture the crossover from the CEL-CEL-p as the starting point, one must include some additional corrections related to the background statistical gauge fields. 
The exact nature of these corrections is currently unclear. 
In fact, there is a theoretical precedence of such extra considerations in the conventional HLR theory, in showing its compatibility with an emergent PH symmetry~\cite{Wang2017,Cheung2017}.  
Although the precise nature of the crossover, especially in the $p$-wave pairing picture, is challenging, it might profit from the duality as discussed in this work. 
This shows the overall importance of our results. 
Namely, upon the transformation, a rather implementable theory emerges, which apart from the fundamental nature of the discussed relations, would aid in further understanding bilayer quantum Hall systems. 
Here, we especially reiterate the natural role of our results in the context of numerical pursuits.

\begin{acknowledgements}
\sect{Acknowledgements} We cordially thank Nigel Cooper and Glenn Wagner for insightful discussions. 
R.-J.~S. and G.~C.~acknowledge funding from a New Investigator Award, EPSRC grant EP/W00187X/1. R.-J.~S also acknowledges funding from Trinity College, University of Cambridge.
L.~R. and R.-J.~S acknowledge funding provided by the Winton programme and the Schiff foundation.
\end{acknowledgements}

\bibliography{bibliography}

\begin{thebibliography}{52}%
\makeatletter
\providecommand \@ifxundefined [1]{%
 \@ifx{#1\undefined}
}%
\providecommand \@ifnum [1]{%
 \ifnum #1\expandafter \@firstoftwo
 \else \expandafter \@secondoftwo
 \fi
}%
\providecommand \@ifx [1]{%
 \ifx #1\expandafter \@firstoftwo
 \else \expandafter \@secondoftwo
 \fi
}%
\providecommand \natexlab [1]{#1}%
\providecommand \enquote  [1]{``#1''}%
\providecommand \bibnamefont  [1]{#1}%
\providecommand \bibfnamefont [1]{#1}%
\providecommand \citenamefont [1]{#1}%
\providecommand \href@noop [0]{\@secondoftwo}%
\providecommand \href [0]{\begingroup \@sanitize@url \@href}%
\providecommand \@href[1]{\@@startlink{#1}\@@href}%
\providecommand \@@href[1]{\endgroup#1\@@endlink}%
\providecommand \@sanitize@url [0]{\catcode `\\12\catcode `\$12\catcode
  `\&12\catcode `\#12\catcode `\^12\catcode `\_12\catcode `\%12\relax}%
\providecommand \@@startlink[1]{}%
\providecommand \@@endlink[0]{}%
\providecommand \url  [0]{\begingroup\@sanitize@url \@url }%
\providecommand \@url [1]{\endgroup\@href {#1}{\urlprefix }}%
\providecommand \urlprefix  [0]{URL }%
\providecommand \Eprint [0]{\href }%
\providecommand \doibase [0]{http://dx.doi.org/}%
\providecommand \selectlanguage [0]{\@gobble}%
\providecommand \bibinfo  [0]{\@secondoftwo}%
\providecommand \bibfield  [0]{\@secondoftwo}%
\providecommand \translation [1]{[#1]}%
\providecommand \BibitemOpen [0]{}%
\providecommand \bibitemStop [0]{}%
\providecommand \bibitemNoStop [0]{.\EOS\space}%
\providecommand \EOS [0]{\spacefactor3000\relax}%
\providecommand \BibitemShut  [1]{\csname bibitem#1\endcsname}%
\let\auto@bib@innerbib\@empty
\bibitem [{\citenamefont {Kramers}\ and\ \citenamefont
  {Wannier}(1941)}]{Kramers1941}%
  \BibitemOpen
  \bibfield  {author} {\bibinfo {author} {\bibfnamefont {H.~A.}\ \bibnamefont
  {Kramers}}\ and\ \bibinfo {author} {\bibfnamefont {G.~H.}\ \bibnamefont
  {Wannier}},\ }\bibfield  {title} {\enquote {\bibinfo {title} {Statistics of
  the two-dimensional ferromagnet. part i},}\ }\href {\doibase
  10.1103/PhysRev.60.252} {\bibfield  {journal} {\bibinfo  {journal} {Phys.
  Rev.}\ }\textbf {\bibinfo {volume} {60}},\ \bibinfo {pages} {252} (\bibinfo
  {year} {1941})}\BibitemShut {NoStop}%
\bibitem [{\citenamefont {Kogut}(1979)}]{Kogut1979}%
  \BibitemOpen
  \bibfield  {author} {\bibinfo {author} {\bibfnamefont {J.~B.}\ \bibnamefont
  {Kogut}},\ }\bibfield  {title} {\enquote {\bibinfo {title} {An introduction
  to lattice gauge theory and spin systems},}\ }\href {\doibase
  10.1103/RevModPhys.51.659} {\bibfield  {journal} {\bibinfo  {journal} {Rev.
  Mod. Phys.}\ }\textbf {\bibinfo {volume} {51}},\ \bibinfo {pages} {659}
  (\bibinfo {year} {1979})}\BibitemShut {NoStop}%
\bibitem [{\citenamefont {Maldacena}(1999)}]{Maldacena_1999}%
  \BibitemOpen
  \bibfield  {author} {\bibinfo {author} {\bibfnamefont {J.}~\bibnamefont
  {Maldacena}},\ }\href {\doibase 10.1023/a:1026654312961} {\bibfield
  {journal} {\bibinfo  {journal} {International Journal of Theoretical
  Physics}\ }\textbf {\bibinfo {volume} {38}},\ \bibinfo {pages} {1113}
  (\bibinfo {year} {1999})}\BibitemShut {NoStop}%
\bibitem [{\citenamefont {Witten}(1998)}]{Witten1998}%
  \BibitemOpen
  \bibfield  {author} {\bibinfo {author} {\bibfnamefont {E.}~\bibnamefont
  {Witten}},\ }\href@noop {} {\enquote {\bibinfo {title} {Anti de sitter space
  and holography},}\ } (\bibinfo {year} {1998}),\ \Eprint
  {http://arxiv.org/abs/hep-th/9802150} {arXiv:hep-th/9802150 [hep-th]}
  \BibitemShut {NoStop}%
\bibitem [{\citenamefont {Beekman}\ \emph {et~al.}(2017)\citenamefont
  {Beekman}, \citenamefont {Nissinen}, \citenamefont {Wu}, \citenamefont {Liu},
  \citenamefont {Slager}, \citenamefont {Nussinov}, \citenamefont {Cvetkovic},\
  and\ \citenamefont {Zaanen}}]{beekman2017}%
  \BibitemOpen
  \bibfield  {author} {\bibinfo {author} {\bibfnamefont {A.~J.}\ \bibnamefont
  {Beekman}}, \bibinfo {author} {\bibfnamefont {J.}~\bibnamefont {Nissinen}},
  \bibinfo {author} {\bibfnamefont {K.}~\bibnamefont {Wu}}, \bibinfo {author}
  {\bibfnamefont {K.}~\bibnamefont {Liu}}, \bibinfo {author} {\bibfnamefont
  {R.-J.}\ \bibnamefont {Slager}}, \bibinfo {author} {\bibfnamefont
  {Z.}~\bibnamefont {Nussinov}}, \bibinfo {author} {\bibfnamefont
  {V.}~\bibnamefont {Cvetkovic}}, \ and\ \bibinfo {author} {\bibfnamefont
  {J.}~\bibnamefont {Zaanen}},\ }\bibfield  {title} {\enquote {\bibinfo {title}
  {Dual gauge field theory of quantum liquid crystals in two dimensions},}\
  }\href {\doibase https://doi.org/10.1016/j.physrep.2017.03.004} {\bibfield
  {journal} {\bibinfo  {journal} {Phys. Rep.}\ }\textbf {\bibinfo {volume}
  {683}},\ \bibinfo {pages} {1} (\bibinfo {year} {2017})}\BibitemShut {NoStop}%
\bibitem [{\citenamefont {Seiberg}\ \emph {et~al.}(2016)\citenamefont
  {Seiberg}, \citenamefont {Senthil}, \citenamefont {Wang},\ and\ \citenamefont
  {Witten}}]{Seiberg2016}%
  \BibitemOpen
  \bibfield  {author} {\bibinfo {author} {\bibfnamefont {N.}~\bibnamefont
  {Seiberg}}, \bibinfo {author} {\bibfnamefont {T.}~\bibnamefont {Senthil}},
  \bibinfo {author} {\bibfnamefont {C.}~\bibnamefont {Wang}}, \ and\ \bibinfo
  {author} {\bibfnamefont {E.}~\bibnamefont {Witten}},\ }\bibfield  {title}
  {\enquote {\bibinfo {title} {A duality web in $2+1$ dimensions and condensed
  matter physics},}\ }\href {\doibase 10.1016/j.aop.2016.08.007} {\bibfield
  {journal} {\bibinfo  {journal} {Ann. Phys.}\ }\textbf {\bibinfo {volume}
  {374}},\ \bibinfo {pages} {395} (\bibinfo {year} {2016})}\BibitemShut
  {NoStop}%
\bibitem [{\citenamefont {Wang}\ and\ \citenamefont
  {Senthil}(2015)}]{Wang2015}%
  \BibitemOpen
  \bibfield  {author} {\bibinfo {author} {\bibfnamefont {C.}~\bibnamefont
  {Wang}}\ and\ \bibinfo {author} {\bibfnamefont {T.}~\bibnamefont {Senthil}},\
  }\bibfield  {title} {\enquote {\bibinfo {title} {Dual dirac liquid on the
  surface of the electron topological insulator},}\ }\href {\doibase
  10.1103/PhysRevX.5.041031} {\bibfield  {journal} {\bibinfo  {journal} {Phys.
  Rev. X}\ }\textbf {\bibinfo {volume} {5}},\ \bibinfo {pages} {041031}
  (\bibinfo {year} {2015})}\BibitemShut {NoStop}%
\bibitem [{\citenamefont {Sodemann}\ \emph {et~al.}(2017)\citenamefont
  {Sodemann}, \citenamefont {Kimchi}, \citenamefont {Wang},\ and\ \citenamefont
  {Senthil}}]{Sodemann2017}%
  \BibitemOpen
  \bibfield  {author} {\bibinfo {author} {\bibfnamefont {I.}~\bibnamefont
  {Sodemann}}, \bibinfo {author} {\bibfnamefont {I.}~\bibnamefont {Kimchi}},
  \bibinfo {author} {\bibfnamefont {C.}~\bibnamefont {Wang}}, \ and\ \bibinfo
  {author} {\bibfnamefont {T.}~\bibnamefont {Senthil}},\ }\bibfield  {title}
  {\enquote {\bibinfo {title} {Composite fermion duality for half-filled
  multicomponent landau levels},}\ }\href {\doibase 10.1103/PhysRevB.95.085135}
  {\bibfield  {journal} {\bibinfo  {journal} {Phys. Rev. B}\ }\textbf {\bibinfo
  {volume} {95}},\ \bibinfo {pages} {085135} (\bibinfo {year}
  {2017})}\BibitemShut {NoStop}%
\bibitem [{\citenamefont {Halperin}(1983)}]{Halperin1983}%
  \BibitemOpen
  \bibfield  {author} {\bibinfo {author} {\bibfnamefont {B.~I.}\ \bibnamefont
  {Halperin}},\ }\bibfield  {title} {\enquote {\bibinfo {title} {Theory of the
  quantized hall conductance},}\ }\href {\doibase 10.5169/seals-115362}
  {\bibfield  {journal} {\bibinfo  {journal} {Helv. Phys. Acta}\ }\textbf
  {\bibinfo {volume} {56}},\ \bibinfo {pages} {75} (\bibinfo {year}
  {1983})}\BibitemShut {NoStop}%
\bibitem [{\citenamefont {Spielman}\ \emph {et~al.}(2000)\citenamefont
  {Spielman}, \citenamefont {Eisenstein}, \citenamefont {Pfeiffer},\ and\
  \citenamefont {West}}]{Spielman2000}%
  \BibitemOpen
  \bibfield  {author} {\bibinfo {author} {\bibfnamefont {I.~B.}\ \bibnamefont
  {Spielman}}, \bibinfo {author} {\bibfnamefont {J.~P.}\ \bibnamefont
  {Eisenstein}}, \bibinfo {author} {\bibfnamefont {L.~N.}\ \bibnamefont
  {Pfeiffer}}, \ and\ \bibinfo {author} {\bibfnamefont {K.~W.}\ \bibnamefont
  {West}},\ }\bibfield  {title} {\enquote {\bibinfo {title} {Resonantly
  enhanced tunneling in a double layer quantum hall ferromagnet},}\ }\href
  {\doibase 10.1103/PhysRevLett.84.5808} {\bibfield  {journal} {\bibinfo
  {journal} {Phys. Rev. Lett.}\ }\textbf {\bibinfo {volume} {84}},\ \bibinfo
  {pages} {5808} (\bibinfo {year} {2000})}\BibitemShut {NoStop}%
\bibitem [{\citenamefont {Eisenstein}\ and\ \citenamefont
  {MacDonald}(2004)}]{Eisenstein2004}%
  \BibitemOpen
  \bibfield  {author} {\bibinfo {author} {\bibfnamefont {J.~P.}\ \bibnamefont
  {Eisenstein}}\ and\ \bibinfo {author} {\bibfnamefont {A.~H.}\ \bibnamefont
  {MacDonald}},\ }\bibfield  {title} {\enquote {\bibinfo {title}
  {Bose--einstein condensation of excitons in bilayer electron systems},}\
  }\href {\doibase 10.1038/nature03081} {\bibfield  {journal} {\bibinfo
  {journal} {Nature}\ }\textbf {\bibinfo {volume} {432}},\ \bibinfo {pages}
  {691} (\bibinfo {year} {2004})}\BibitemShut {NoStop}%
\bibitem [{\citenamefont {Halperin}\ \emph {et~al.}(1993)\citenamefont
  {Halperin}, \citenamefont {Lee},\ and\ \citenamefont {Read}}]{Halperin1993}%
  \BibitemOpen
  \bibfield  {author} {\bibinfo {author} {\bibfnamefont {B.~I.}\ \bibnamefont
  {Halperin}}, \bibinfo {author} {\bibfnamefont {Patrick~A.}\ \bibnamefont
  {Lee}}, \ and\ \bibinfo {author} {\bibfnamefont {Nicholas}\ \bibnamefont
  {Read}},\ }\bibfield  {title} {\enquote {\bibinfo {title} {Theory of the
  half-filled landau level},}\ }\href {\doibase 10.1103/PhysRevB.47.7312}
  {\bibfield  {journal} {\bibinfo  {journal} {Phys. Rev. B}\ }\textbf {\bibinfo
  {volume} {47}},\ \bibinfo {pages} {7312} (\bibinfo {year}
  {1993})}\BibitemShut {NoStop}%
\bibitem [{\citenamefont {Bonesteel}(1993)}]{Bonesteel1993}%
  \BibitemOpen
  \bibfield  {author} {\bibinfo {author} {\bibfnamefont {N.~E.}\ \bibnamefont
  {Bonesteel}},\ }\bibfield  {title} {\enquote {\bibinfo {title} {Compressible
  phase of a double-layer electron system with total landau-level filling
  factor 1/2},}\ }\href {\doibase 10.1103/PhysRevB.48.11484} {\bibfield
  {journal} {\bibinfo  {journal} {Phys. Rev. B}\ }\textbf {\bibinfo {volume}
  {48}},\ \bibinfo {pages} {11484} (\bibinfo {year} {1993})}\BibitemShut
  {NoStop}%
\bibitem [{\citenamefont {Bonesteel}\ \emph {et~al.}(1996)\citenamefont
  {Bonesteel}, \citenamefont {McDonald},\ and\ \citenamefont
  {Nayak}}]{Bonesteel1996}%
  \BibitemOpen
  \bibfield  {author} {\bibinfo {author} {\bibfnamefont {N.~E.}\ \bibnamefont
  {Bonesteel}}, \bibinfo {author} {\bibfnamefont {I.~A.}\ \bibnamefont
  {McDonald}}, \ and\ \bibinfo {author} {\bibfnamefont {C.}~\bibnamefont
  {Nayak}},\ }\bibfield  {title} {\enquote {\bibinfo {title} {Gauge fields and
  pairing in double-layer composite fermion metals},}\ }\href {\doibase
  10.1103/PhysRevLett.77.3009} {\bibfield  {journal} {\bibinfo  {journal}
  {Phys. Rev. Lett.}\ }\textbf {\bibinfo {volume} {77}},\ \bibinfo {pages}
  {3009} (\bibinfo {year} {1996})}\BibitemShut {NoStop}%
\bibitem [{\citenamefont {Morinari}(1999)}]{Morinari1999}%
  \BibitemOpen
  \bibfield  {author} {\bibinfo {author} {\bibfnamefont {T.}~\bibnamefont
  {Morinari}},\ }\bibfield  {title} {\enquote {\bibinfo {title}
  {Composite-fermion pairing in bilayer quantum hall systems},}\ }\href
  {\doibase 10.1103/PhysRevB.59.7320} {\bibfield  {journal} {\bibinfo
  {journal} {Phys. Rev. B}\ }\textbf {\bibinfo {volume} {59}},\ \bibinfo
  {pages} {7320} (\bibinfo {year} {1999})}\BibitemShut {NoStop}%
\bibitem [{\citenamefont {Kim}\ \emph {et~al.}(2001)\citenamefont {Kim},
  \citenamefont {Nayak}, \citenamefont {Demler}, \citenamefont {Read},\ and\
  \citenamefont {Das~Sarma}}]{Kim2001}%
  \BibitemOpen
  \bibfield  {author} {\bibinfo {author} {\bibfnamefont {Y.~B.}\ \bibnamefont
  {Kim}}, \bibinfo {author} {\bibfnamefont {C.}~\bibnamefont {Nayak}}, \bibinfo
  {author} {\bibfnamefont {E.}~\bibnamefont {Demler}}, \bibinfo {author}
  {\bibfnamefont {N.}~\bibnamefont {Read}}, \ and\ \bibinfo {author}
  {\bibfnamefont {S.}~\bibnamefont {Das~Sarma}},\ }\bibfield  {title} {\enquote
  {\bibinfo {title} {Bilayer paired quantum hall states and coulomb drag},}\
  }\href {\doibase 10.1103/PhysRevB.63.205315} {\bibfield  {journal} {\bibinfo
  {journal} {Phys. Rev. B}\ }\textbf {\bibinfo {volume} {63}},\ \bibinfo
  {pages} {205315} (\bibinfo {year} {2001})}\BibitemShut {NoStop}%
\bibitem [{\citenamefont {Simon}\ \emph {et~al.}(2003)\citenamefont {Simon},
  \citenamefont {Rezayi},\ and\ \citenamefont {Milovanovic}}]{Simon2003}%
  \BibitemOpen
  \bibfield  {author} {\bibinfo {author} {\bibfnamefont {S.~H.}\ \bibnamefont
  {Simon}}, \bibinfo {author} {\bibfnamefont {E.~H.}\ \bibnamefont {Rezayi}}, \
  and\ \bibinfo {author} {\bibfnamefont {M.~V.}\ \bibnamefont {Milovanovic}},\
  }\bibfield  {title} {\enquote {\bibinfo {title} {Coexistence of composite
  bosons and composite fermions in $\ensuremath{\nu}=\frac{1}{2}+\frac{1}{2}$
  quantum hall bilayers},}\ }\href {\doibase 10.1103/PhysRevLett.91.046803}
  {\bibfield  {journal} {\bibinfo  {journal} {Phys. Rev. Lett.}\ }\textbf
  {\bibinfo {volume} {91}},\ \bibinfo {pages} {046803} (\bibinfo {year}
  {2003})}\BibitemShut {NoStop}%
\bibitem [{\citenamefont {Shibata}\ and\ \citenamefont
  {Yoshioka}(2006)}]{Shibata2006}%
  \BibitemOpen
  \bibfield  {author} {\bibinfo {author} {\bibfnamefont {N.}~\bibnamefont
  {Shibata}}\ and\ \bibinfo {author} {\bibfnamefont {D.}~\bibnamefont
  {Yoshioka}},\ }\bibfield  {title} {\enquote {\bibinfo {title} {Ground state
  of $\nu=1$ bilayer quantum hall systems},}\ }\href {\doibase
  10.1143/JPSJ.75.043712} {\bibfield  {journal} {\bibinfo  {journal} {JPSJ}\
  }\textbf {\bibinfo {volume} {75}},\ \bibinfo {pages} {043712} (\bibinfo
  {year} {2006})}\BibitemShut {NoStop}%
\bibitem [{\citenamefont {Doretto}\ \emph {et~al.}(2006)\citenamefont
  {Doretto}, \citenamefont {Caldeira},\ and\ \citenamefont
  {Smith}}]{Doretto2006}%
  \BibitemOpen
  \bibfield  {author} {\bibinfo {author} {\bibfnamefont {R.~L.}\ \bibnamefont
  {Doretto}}, \bibinfo {author} {\bibfnamefont {A.~O.}\ \bibnamefont
  {Caldeira}}, \ and\ \bibinfo {author} {\bibfnamefont {M.~C.}\ \bibnamefont
  {Smith}},\ }\bibfield  {title} {\enquote {\bibinfo {title} {Bosonization
  approach for bilayer quantum hall systems at ${\ensuremath{\nu}}_{T}=1$},}\
  }\href {\doibase 10.1103/PhysRevLett.97.186401} {\bibfield  {journal}
  {\bibinfo  {journal} {Phys. Rev. Lett.}\ }\textbf {\bibinfo {volume} {97}},\
  \bibinfo {pages} {186401} (\bibinfo {year} {2006})}\BibitemShut {NoStop}%
\bibitem [{\citenamefont {Zhu}\ \emph {et~al.}(2017)\citenamefont {Zhu},
  \citenamefont {Fu},\ and\ \citenamefont {Sheng}}]{Zhu2017}%
  \BibitemOpen
  \bibfield  {author} {\bibinfo {author} {\bibfnamefont {Z.}~\bibnamefont
  {Zhu}}, \bibinfo {author} {\bibfnamefont {L.}~\bibnamefont {Fu}}, \ and\
  \bibinfo {author} {\bibfnamefont {D.~N.}\ \bibnamefont {Sheng}},\ }\bibfield
  {title} {\enquote {\bibinfo {title} {Numerical study of quantum hall bilayers
  at total filling ${\ensuremath{\nu}}_{T}=1$: A new phase at intermediate
  layer distances},}\ }\href {\doibase 10.1103/PhysRevLett.119.177601}
  {\bibfield  {journal} {\bibinfo  {journal} {Phys. Rev. Lett.}\ }\textbf
  {\bibinfo {volume} {119}},\ \bibinfo {pages} {177601} (\bibinfo {year}
  {2017})}\BibitemShut {NoStop}%
\bibitem [{\citenamefont {Lian}\ and\ \citenamefont {Zhang}(2018)}]{Lian2018}%
  \BibitemOpen
  \bibfield  {author} {\bibinfo {author} {\bibfnamefont {B.}~\bibnamefont
  {Lian}}\ and\ \bibinfo {author} {\bibfnamefont {S.-C.}\ \bibnamefont
  {Zhang}},\ }\bibfield  {title} {\enquote {\bibinfo {title} {Wave function and
  emergent su(2) symmetry in the ${\ensuremath{\nu}}_{T}=1$ quantum hall
  bilayer},}\ }\href {\doibase 10.1103/PhysRevLett.120.077601} {\bibfield
  {journal} {\bibinfo  {journal} {Phys. Rev. Lett.}\ }\textbf {\bibinfo
  {volume} {120}},\ \bibinfo {pages} {077601} (\bibinfo {year}
  {2018})}\BibitemShut {NoStop}%
\bibitem [{\citenamefont {Cipri}\ and\ \citenamefont
  {Bonesteel}(2014)}]{Cipri2014}%
  \BibitemOpen
  \bibfield  {author} {\bibinfo {author} {\bibfnamefont {R.}~\bibnamefont
  {Cipri}}\ and\ \bibinfo {author} {\bibfnamefont {N.~E.}\ \bibnamefont
  {Bonesteel}},\ }\bibfield  {title} {\enquote {\bibinfo {title} {Gauge
  fluctuations and interlayer coherence in bilayer composite fermion metals},}\
  }\href {\doibase 10.1103/PhysRevB.89.085109} {\bibfield  {journal} {\bibinfo
  {journal} {Phys. Rev. B}\ }\textbf {\bibinfo {volume} {89}},\ \bibinfo
  {pages} {085109} (\bibinfo {year} {2014})}\BibitemShut {NoStop}%
\bibitem [{\citenamefont {Isobe}\ and\ \citenamefont {Fu}(2017)}]{Isobe2017}%
  \BibitemOpen
  \bibfield  {author} {\bibinfo {author} {\bibfnamefont {H.}~\bibnamefont
  {Isobe}}\ and\ \bibinfo {author} {\bibfnamefont {L.}~\bibnamefont {Fu}},\
  }\bibfield  {title} {\enquote {\bibinfo {title} {Interlayer pairing symmetry
  of composite fermions in quantum hall bilayers},}\ }\href {\doibase
  10.1103/PhysRevLett.118.166401} {\bibfield  {journal} {\bibinfo  {journal}
  {Phys. Rev. Lett.}\ }\textbf {\bibinfo {volume} {118}},\ \bibinfo {pages}
  {166401} (\bibinfo {year} {2017})}\BibitemShut {NoStop}%
\bibitem [{\citenamefont {M\"oller}\ \emph {et~al.}(2008)\citenamefont
  {M\"oller}, \citenamefont {Simon},\ and\ \citenamefont
  {Rezayi}}]{Moller2008}%
  \BibitemOpen
  \bibfield  {author} {\bibinfo {author} {\bibfnamefont {G.}~\bibnamefont
  {M\"oller}}, \bibinfo {author} {\bibfnamefont {S.~H.}\ \bibnamefont {Simon}},
  \ and\ \bibinfo {author} {\bibfnamefont {E.~H.}\ \bibnamefont {Rezayi}},\
  }\bibfield  {title} {\enquote {\bibinfo {title} {Paired composite fermion
  phase of quantum hall bilayers at
  $\ensuremath{\nu}=\frac{1}{2}+\frac{1}{2}$},}\ }\href {\doibase
  10.1103/PhysRevLett.101.176803} {\bibfield  {journal} {\bibinfo  {journal}
  {Phys. Rev. Lett.}\ }\textbf {\bibinfo {volume} {101}},\ \bibinfo {pages}
  {176803} (\bibinfo {year} {2008})}\BibitemShut {NoStop}%
\bibitem [{\citenamefont {M\"oller}\ \emph {et~al.}(2009)\citenamefont
  {M\"oller}, \citenamefont {Simon},\ and\ \citenamefont
  {Rezayi}}]{Moller2009}%
  \BibitemOpen
  \bibfield  {author} {\bibinfo {author} {\bibfnamefont {G.}~\bibnamefont
  {M\"oller}}, \bibinfo {author} {\bibfnamefont {S.~H.}\ \bibnamefont {Simon}},
  \ and\ \bibinfo {author} {\bibfnamefont {E.~H.}\ \bibnamefont {Rezayi}},\
  }\bibfield  {title} {\enquote {\bibinfo {title} {Trial wave functions for
  $\ensuremath{\nu}=\frac{1}{2}+\frac{1}{2}$ quantum hall bilayers},}\ }\href
  {\doibase 10.1103/PhysRevB.79.125106} {\bibfield  {journal} {\bibinfo
  {journal} {Phys. Rev. B}\ }\textbf {\bibinfo {volume} {79}},\ \bibinfo
  {pages} {125106} (\bibinfo {year} {2009})}\BibitemShut {NoStop}%
\bibitem [{Note1()}]{Note1}%
  \BibitemOpen
  \bibinfo {note} {In the Supplemental Material, we show that the dominant
  channel rather corresponds to $l = -1$. This is, however, only a semantical
  distinction and we agree with the results of the aforementioned
  references.}\BibitemShut {Stop}%
\bibitem [{\citenamefont {Liu}\ \emph {et~al.}(2022)\citenamefont {Liu},
  \citenamefont {Li}, \citenamefont {Watanabe}, \citenamefont {Taniguchi},
  \citenamefont {Hone}, \citenamefont {Halperin}, \citenamefont {Kim},\ and\
  \citenamefont {Dean}}]{Liu2022}%
  \BibitemOpen
  \bibfield  {author} {\bibinfo {author} {\bibfnamefont {X.}~\bibnamefont
  {Liu}}, \bibinfo {author} {\bibfnamefont {J.~I.~A.}\ \bibnamefont {Li}},
  \bibinfo {author} {\bibfnamefont {K.}~\bibnamefont {Watanabe}}, \bibinfo
  {author} {\bibfnamefont {T.}~\bibnamefont {Taniguchi}}, \bibinfo {author}
  {\bibfnamefont {J.}~\bibnamefont {Hone}}, \bibinfo {author} {\bibfnamefont
  {B.~I.}\ \bibnamefont {Halperin}}, \bibinfo {author} {\bibfnamefont
  {P.}~\bibnamefont {Kim}}, \ and\ \bibinfo {author} {\bibfnamefont {C.~R.}\
  \bibnamefont {Dean}},\ }\bibfield  {title} {\enquote {\bibinfo {title}
  {Crossover between strongly coupled and weakly coupled exciton
  superfluids},}\ }\href {\doibase 10.1126/science.abg1110} {\bibfield
  {journal} {\bibinfo  {journal} {Science}\ }\textbf {\bibinfo {volume}
  {375}},\ \bibinfo {pages} {205} (\bibinfo {year} {2022})}\BibitemShut
  {NoStop}%
\bibitem [{\citenamefont {Eisenstein}\ \emph {et~al.}(2019)\citenamefont
  {Eisenstein}, \citenamefont {Pfeiffer},\ and\ \citenamefont
  {West}}]{Eisenstein2019}%
  \BibitemOpen
  \bibfield  {author} {\bibinfo {author} {\bibfnamefont {J.~P.}\ \bibnamefont
  {Eisenstein}}, \bibinfo {author} {\bibfnamefont {L.~N.}\ \bibnamefont
  {Pfeiffer}}, \ and\ \bibinfo {author} {\bibfnamefont {K.~W.}\ \bibnamefont
  {West}},\ }\bibfield  {title} {\enquote {\bibinfo {title} {Precursors to
  exciton condensation in quantum hall bilayers},}\ }\href {\doibase
  10.1103/PhysRevLett.123.066802} {\bibfield  {journal} {\bibinfo  {journal}
  {Phys. Rev. Lett.}\ }\textbf {\bibinfo {volume} {123}},\ \bibinfo {pages}
  {066802} (\bibinfo {year} {2019})}\BibitemShut {NoStop}%
\bibitem [{\citenamefont {Wagner}\ \emph {et~al.}(2021)\citenamefont {Wagner},
  \citenamefont {Nguyen}, \citenamefont {Simon},\ and\ \citenamefont
  {Halperin}}]{Wagner2021}%
  \BibitemOpen
  \bibfield  {author} {\bibinfo {author} {\bibfnamefont {G.}~\bibnamefont
  {Wagner}}, \bibinfo {author} {\bibfnamefont {D.~X.}\ \bibnamefont {Nguyen}},
  \bibinfo {author} {\bibfnamefont {S.~H.}\ \bibnamefont {Simon}}, \ and\
  \bibinfo {author} {\bibfnamefont {B.~I.}\ \bibnamefont {Halperin}},\
  }\bibfield  {title} {\enquote {\bibinfo {title} {$s$-wave paired electron and
  hole composite fermion trial state for quantum hall bilayers with
  $\ensuremath{\nu}=1$},}\ }\href {\doibase 10.1103/PhysRevLett.127.246803}
  {\bibfield  {journal} {\bibinfo  {journal} {Phys. Rev. Lett.}\ }\textbf
  {\bibinfo {volume} {127}},\ \bibinfo {pages} {246803} (\bibinfo {year}
  {2021})}\BibitemShut {NoStop}%
\bibitem [{\citenamefont {R\"uegg}\ \emph {et~al.}(2023)\citenamefont
  {R\"uegg}, \citenamefont {Chaudhary},\ and\ \citenamefont
  {Slager}}]{Rueegg2023}%
  \BibitemOpen
  \bibfield  {author} {\bibinfo {author} {\bibfnamefont {L.}~\bibnamefont
  {R\"uegg}}, \bibinfo {author} {\bibfnamefont {G.}~\bibnamefont {Chaudhary}},
  \ and\ \bibinfo {author} {\bibfnamefont {R.-J.}\ \bibnamefont {Slager}},\
  }\bibfield  {title} {\enquote {\bibinfo {title} {Pairing of composite
  electrons and composite holes in ${\ensuremath{\nu}}_{T}=1$ quantum hall
  bilayers},}\ }\href {\doibase 10.1103/PhysRevResearch.5.L042022} {\bibfield
  {journal} {\bibinfo  {journal} {Phys. Rev. Res.}\ }\textbf {\bibinfo {volume}
  {5}},\ \bibinfo {pages} {L042022} (\bibinfo {year} {2023})}\BibitemShut
  {NoStop}%
\bibitem [{\citenamefont {Predin}\ \emph {et~al.}(2023)\citenamefont {Predin},
  \citenamefont {Kne\ifmmode \check{z}\else
  \v{z}\fi{}evi\ifmmode~\acute{c}\else \'{c}\fi{}},\ and\ \citenamefont
  {Milovanovi\ifmmode~\acute{c}\else \'{c}\fi{}}}]{Predin2023}%
  \BibitemOpen
  \bibfield  {author} {\bibinfo {author} {\bibfnamefont {S.}~\bibnamefont
  {Predin}}, \bibinfo {author} {\bibfnamefont {A.}~\bibnamefont {Kne\ifmmode
  \check{z}\else \v{z}\fi{}evi\ifmmode~\acute{c}\else \'{c}\fi{}}}, \ and\
  \bibinfo {author} {\bibfnamefont {M.~V.}\ \bibnamefont
  {Milovanovi\ifmmode~\acute{c}\else \'{c}\fi{}}},\ }\bibfield  {title}
  {\enquote {\bibinfo {title} {Dipole representation of half-filled landau
  level},}\ }\href {\doibase 10.1103/PhysRevB.107.155132} {\bibfield  {journal}
  {\bibinfo  {journal} {Phys. Rev. B}\ }\textbf {\bibinfo {volume} {107}},\
  \bibinfo {pages} {155132} (\bibinfo {year} {2023})}\BibitemShut {NoStop}%
\bibitem [{\citenamefont {Predin}\ and\ \citenamefont
  {Milovanovi\ifmmode~\acute{c}\else \'{c}\fi{}}(2023)}]{Predin2023a}%
  \BibitemOpen
  \bibfield  {author} {\bibinfo {author} {\bibfnamefont {S.}~\bibnamefont
  {Predin}}\ and\ \bibinfo {author} {\bibfnamefont {M.~V.}\ \bibnamefont
  {Milovanovi\ifmmode~\acute{c}\else \'{c}\fi{}}},\ }\bibfield  {title}
  {\enquote {\bibinfo {title} {Quantum hall bilayer in dipole
  representation},}\ }\href {\doibase 10.1103/PhysRevB.108.155129} {\bibfield
  {journal} {\bibinfo  {journal} {Phys. Rev. B}\ }\textbf {\bibinfo {volume}
  {108}},\ \bibinfo {pages} {155129} (\bibinfo {year} {2023})}\BibitemShut
  {NoStop}%
\bibitem [{\citenamefont {Tev\v{z}}\ and\ \citenamefont
  {Simon}(2024)}]{Lotric2024}%
  \BibitemOpen
  \bibfield  {author} {\bibinfo {author} {\bibfnamefont {L.}~\bibnamefont
  {Tev\v{z}}}\ and\ \bibinfo {author} {\bibfnamefont {S.~H.}\ \bibnamefont
  {Simon}},\ }\bibfield  {title} {\enquote {\bibinfo {title} {Chern-simons
  modified rpa-eliashberg theory of the
  $\ensuremath{\nu}=\frac{1}{2}+\frac{1}{2}$ quantum hall bilayer},}\ }\href
  {\doibase 10.1103/PhysRevLett.132.176502} {\bibfield  {journal} {\bibinfo
  {journal} {Phys. Rev. Lett.}\ }\textbf {\bibinfo {volume} {132}},\ \bibinfo
  {pages} {176502} (\bibinfo {year} {2024})}\BibitemShut {NoStop}%
\bibitem [{\citenamefont {Son}(2015)}]{Son2015}%
  \BibitemOpen
  \bibfield  {author} {\bibinfo {author} {\bibfnamefont {D.~T.}\ \bibnamefont
  {Son}},\ }\bibfield  {title} {\enquote {\bibinfo {title} {Is the composite
  fermion a dirac particle?}}\ }\href {\doibase 10.1103/PhysRevX.5.031027}
  {\bibfield  {journal} {\bibinfo  {journal} {Phys. Rev. X}\ }\textbf {\bibinfo
  {volume} {5}},\ \bibinfo {pages} {031027} (\bibinfo {year}
  {2015})}\BibitemShut {NoStop}%
\bibitem [{\citenamefont {Wang}\ \emph {et~al.}(2017)\citenamefont {Wang},
  \citenamefont {Cooper}, \citenamefont {Halperin},\ and\ \citenamefont
  {Stern}}]{Wang2017}%
  \BibitemOpen
  \bibfield  {author} {\bibinfo {author} {\bibfnamefont {C.}~\bibnamefont
  {Wang}}, \bibinfo {author} {\bibfnamefont {N.~R.}\ \bibnamefont {Cooper}},
  \bibinfo {author} {\bibfnamefont {B.~I.}\ \bibnamefont {Halperin}}, \ and\
  \bibinfo {author} {\bibfnamefont {A.}~\bibnamefont {Stern}},\ }\bibfield
  {title} {\enquote {\bibinfo {title} {Particle-hole symmetry in the
  fermion-chern-simons and dirac descriptions of a half-filled landau level},}\
  }\href {\doibase 10.1103/PhysRevX.7.031029} {\bibfield  {journal} {\bibinfo
  {journal} {Phys. Rev. X}\ }\textbf {\bibinfo {volume} {7}},\ \bibinfo {pages}
  {031029} (\bibinfo {year} {2017})}\BibitemShut {NoStop}%
\bibitem [{\citenamefont {Cheung}\ \emph {et~al.}(2017)\citenamefont {Cheung},
  \citenamefont {Raghu},\ and\ \citenamefont {Mulligan}}]{Cheung2017}%
  \BibitemOpen
  \bibfield  {author} {\bibinfo {author} {\bibfnamefont {A.~K.~C.}\
  \bibnamefont {Cheung}}, \bibinfo {author} {\bibfnamefont {S.}~\bibnamefont
  {Raghu}}, \ and\ \bibinfo {author} {\bibfnamefont {M.}~\bibnamefont
  {Mulligan}},\ }\bibfield  {title} {\enquote {\bibinfo {title} {Weiss
  oscillations and particle-hole symmetry at the half-filled landau level},}\
  }\href {\doibase 10.1103/PhysRevB.95.235424} {\bibfield  {journal} {\bibinfo
  {journal} {Phys. Rev. B}\ }\textbf {\bibinfo {volume} {95}},\ \bibinfo
  {pages} {235424} (\bibinfo {year} {2017})}\BibitemShut {NoStop}%
\bibitem [{\citenamefont {Wen}\ and\ \citenamefont {Zee}(1992)}]{Wen1992}%
  \BibitemOpen
  \bibfield  {author} {\bibinfo {author} {\bibfnamefont {X.-G.}\ \bibnamefont
  {Wen}}\ and\ \bibinfo {author} {\bibfnamefont {A.}~\bibnamefont {Zee}},\
  }\bibfield  {title} {\enquote {\bibinfo {title} {Neutral superfluid modes and
  ``magnetic'' monopoles in multilayered quantum hall systems},}\ }\href
  {\doibase 10.1103/PhysRevLett.69.1811} {\bibfield  {journal} {\bibinfo
  {journal} {Phys. Rev. Lett.}\ }\textbf {\bibinfo {volume} {69}},\ \bibinfo
  {pages} {1811} (\bibinfo {year} {1992})}\BibitemShut {NoStop}%
\bibitem [{\citenamefont {Wen}\ and\ \citenamefont {Zee}(1993)}]{Wen1993}%
  \BibitemOpen
  \bibfield  {author} {\bibinfo {author} {\bibfnamefont {X.-G.}\ \bibnamefont
  {Wen}}\ and\ \bibinfo {author} {\bibfnamefont {A.}~\bibnamefont {Zee}},\
  }\bibfield  {title} {\enquote {\bibinfo {title} {Tunneling in double-layered
  quantum hall systems},}\ }\href {\doibase 10.1103/PhysRevB.47.2265}
  {\bibfield  {journal} {\bibinfo  {journal} {Phys. Rev. B}\ }\textbf {\bibinfo
  {volume} {47}},\ \bibinfo {pages} {2265} (\bibinfo {year}
  {1993})}\BibitemShut {NoStop}%
\bibitem [{\citenamefont {Barkeshli}\ \emph {et~al.}(2015)\citenamefont
  {Barkeshli}, \citenamefont {Mulligan},\ and\ \citenamefont
  {Fisher}}]{Barkeshli2015}%
  \BibitemOpen
  \bibfield  {author} {\bibinfo {author} {\bibfnamefont {M.}~\bibnamefont
  {Barkeshli}}, \bibinfo {author} {\bibfnamefont {M.}~\bibnamefont {Mulligan}},
  \ and\ \bibinfo {author} {\bibfnamefont {M.~P.~A.}\ \bibnamefont {Fisher}},\
  }\bibfield  {title} {\enquote {\bibinfo {title} {Particle-hole symmetry and
  the composite fermi liquid},}\ }\href {\doibase 10.1103/PhysRevB.92.165125}
  {\bibfield  {journal} {\bibinfo  {journal} {Phys. Rev. B}\ }\textbf {\bibinfo
  {volume} {92}},\ \bibinfo {pages} {165125} (\bibinfo {year}
  {2015})}\BibitemShut {NoStop}%
\bibitem [{\citenamefont {Girvin}(1984)}]{Girvin1984}%
  \BibitemOpen
  \bibfield  {author} {\bibinfo {author} {\bibfnamefont {S.~M.}\ \bibnamefont
  {Girvin}},\ }\bibfield  {title} {\enquote {\bibinfo {title} {Particle-hole
  symmetry in the anomalous quantum hall effect},}\ }\href {\doibase
  10.1103/PhysRevB.29.6012} {\bibfield  {journal} {\bibinfo  {journal} {Phys.
  Rev. B}\ }\textbf {\bibinfo {volume} {29}},\ \bibinfo {pages} {6012}
  (\bibinfo {year} {1984})}\BibitemShut {NoStop}%
\bibitem [{\citenamefont {Kivelson}\ \emph {et~al.}(1997)\citenamefont
  {Kivelson}, \citenamefont {Lee}, \citenamefont {Krotov},\ and\ \citenamefont
  {Gan}}]{Kivelson1997}%
  \BibitemOpen
  \bibfield  {author} {\bibinfo {author} {\bibfnamefont {S.~A.}\ \bibnamefont
  {Kivelson}}, \bibinfo {author} {\bibfnamefont {D-H.}\ \bibnamefont {Lee}},
  \bibinfo {author} {\bibfnamefont {Y.}~\bibnamefont {Krotov}}, \ and\ \bibinfo
  {author} {\bibfnamefont {J.}~\bibnamefont {Gan}},\ }\bibfield  {title}
  {\enquote {\bibinfo {title} {Composite-fermion hall conductance at
  \ensuremath{\nu}=},}\ }\href {\doibase 10.1103/PhysRevB.55.15552} {\bibfield
  {journal} {\bibinfo  {journal} {Phys. Rev. B}\ }\textbf {\bibinfo {volume}
  {55}},\ \bibinfo {pages} {15552} (\bibinfo {year} {1997})}\BibitemShut
  {NoStop}%
\bibitem [{\citenamefont {Rezayi}\ and\ \citenamefont
  {Haldane}(2000)}]{Rezayi2000}%
  \BibitemOpen
  \bibfield  {author} {\bibinfo {author} {\bibfnamefont {E.~H.}\ \bibnamefont
  {Rezayi}}\ and\ \bibinfo {author} {\bibfnamefont {F.~D.~M.}\ \bibnamefont
  {Haldane}},\ }\bibfield  {title} {\enquote {\bibinfo {title} {Incompressible
  paired hall state, stripe order, and the composite fermion liquid phase in
  half-filled landau levels},}\ }\href {\doibase 10.1103/PhysRevLett.84.4685}
  {\bibfield  {journal} {\bibinfo  {journal} {Phys. Rev. Lett.}\ }\textbf
  {\bibinfo {volume} {84}},\ \bibinfo {pages} {4685} (\bibinfo {year}
  {2000})}\BibitemShut {NoStop}%
\bibitem [{\citenamefont {Balram}\ and\ \citenamefont
  {Jain}(2016)}]{Balram2016}%
  \BibitemOpen
  \bibfield  {author} {\bibinfo {author} {\bibfnamefont {A.~C.}\ \bibnamefont
  {Balram}}\ and\ \bibinfo {author} {\bibfnamefont {J.~K.}\ \bibnamefont
  {Jain}},\ }\bibfield  {title} {\enquote {\bibinfo {title} {Nature of
  composite fermions and the role of particle-hole symmetry: A microscopic
  account},}\ }\href {\doibase 10.1103/PhysRevB.93.235152} {\bibfield
  {journal} {\bibinfo  {journal} {Phys. Rev. B}\ }\textbf {\bibinfo {volume}
  {93}},\ \bibinfo {pages} {235152} (\bibinfo {year} {2016})}\BibitemShut
  {NoStop}%
\bibitem [{\citenamefont {Geraedts}\ \emph {et~al.}(2016)\citenamefont
  {Geraedts}, \citenamefont {Zaletel}, \citenamefont {Mong}, \citenamefont
  {Metlitski}, \citenamefont {Vishwanath},\ and\ \citenamefont
  {Motrunich}}]{Geraedts2016}%
  \BibitemOpen
  \bibfield  {author} {\bibinfo {author} {\bibfnamefont {S.~D.}\ \bibnamefont
  {Geraedts}}, \bibinfo {author} {\bibfnamefont {M.~P.}\ \bibnamefont
  {Zaletel}}, \bibinfo {author} {\bibfnamefont {R.~S.~K.}\ \bibnamefont
  {Mong}}, \bibinfo {author} {\bibfnamefont {M.~A.}\ \bibnamefont {Metlitski}},
  \bibinfo {author} {\bibfnamefont {A.}~\bibnamefont {Vishwanath}}, \ and\
  \bibinfo {author} {\bibfnamefont {O.~I.}\ \bibnamefont {Motrunich}},\
  }\bibfield  {title} {\enquote {\bibinfo {title} {The half-filled landau
  level: The case for dirac composite fermions},}\ }\href {\doibase
  10.1126/science.aad4302} {\bibfield  {journal} {\bibinfo  {journal}
  {Science}\ }\textbf {\bibinfo {volume} {352}},\ \bibinfo {pages} {197}
  (\bibinfo {year} {2016})}\BibitemShut {NoStop}%
\bibitem [{sup()}]{supplementary}%
  \BibitemOpen
  \href@noop {} {\bibinfo  {journal} {See Supplemental Material for arguments
  to settle the discussion about the sign of the pairing channel, how the
  composite hole liquid theory can be derived from the particle-hole transform
  of a Dirac composite fermion theory, the details of the Bogliubov-de Gennes
  calculations, and a thorough discussion of the wave function arguments.}\
  }\BibitemShut {NoStop}%
\bibitem [{\citenamefont {Fu}\ and\ \citenamefont {Kane}(2008)}]{Fu2008}%
  \BibitemOpen
\bibfield  {journal} {  }\bibfield  {author} {\bibinfo {author} {\bibfnamefont
  {L.}~\bibnamefont {Fu}}\ and\ \bibinfo {author} {\bibfnamefont {C.~L.}\
  \bibnamefont {Kane}},\ }\bibfield  {title} {\enquote {\bibinfo {title}
  {Superconducting proximity effect and majorana fermions at the surface of a
  topological insulator},}\ }\href {\doibase 10.1103/PhysRevLett.100.096407}
  {\bibfield  {journal} {\bibinfo  {journal} {Phys. Rev. Lett.}\ }\textbf
  {\bibinfo {volume} {100}},\ \bibinfo {pages} {096407} (\bibinfo {year}
  {2008})}\BibitemShut {NoStop}%
\bibitem [{\citenamefont {Hansson}\ \emph {et~al.}(2004)\citenamefont
  {Hansson}, \citenamefont {Oganesyan},\ and\ \citenamefont
  {Sondhi}}]{Hansson2004}%
  \BibitemOpen
  \bibfield  {author} {\bibinfo {author} {\bibfnamefont {T.~H.}\ \bibnamefont
  {Hansson}}, \bibinfo {author} {\bibfnamefont {V.}~\bibnamefont {Oganesyan}},
  \ and\ \bibinfo {author} {\bibfnamefont {S.~L.}\ \bibnamefont {Sondhi}},\
  }\bibfield  {title} {\enquote {\bibinfo {title} {Superconductors are
  topologically ordered},}\ }\href {\doibase
  https://doi.org/10.1016/j.aop.2004.05.006} {\bibfield  {journal} {\bibinfo
  {journal} {Ann. Phys.}\ }\textbf {\bibinfo {volume} {313}},\ \bibinfo {pages}
  {497} (\bibinfo {year} {2004})}\BibitemShut {NoStop}%
\bibitem [{\citenamefont {Kivelson}\ and\ \citenamefont
  {Rokhsar}(1990)}]{Kivelson1990}%
  \BibitemOpen
  \bibfield  {author} {\bibinfo {author} {\bibfnamefont {S.~A.}\ \bibnamefont
  {Kivelson}}\ and\ \bibinfo {author} {\bibfnamefont {D.~S.}\ \bibnamefont
  {Rokhsar}},\ }\bibfield  {title} {\enquote {\bibinfo {title} {Bogoliubov
  quasiparticles, spinons, and spin-charge decoupling in superconductors},}\
  }\href {\doibase 10.1103/PhysRevB.41.11693} {\bibfield  {journal} {\bibinfo
  {journal} {Phys. Rev. B}\ }\textbf {\bibinfo {volume} {41}},\ \bibinfo
  {pages} {11693} (\bibinfo {year} {1990})}\BibitemShut {NoStop}%
\bibitem [{\citenamefont {Blok}\ and\ \citenamefont {Wen}(1990)}]{Blok1990}%
  \BibitemOpen
  \bibfield  {author} {\bibinfo {author} {\bibfnamefont {B.}~\bibnamefont
  {Blok}}\ and\ \bibinfo {author} {\bibfnamefont {X.~G.}\ \bibnamefont {Wen}},\
  }\bibfield  {title} {\enquote {\bibinfo {title} {Effective theories of the
  fractional quantum hall effect at generic filling fractions},}\ }\href
  {\doibase 10.1103/PhysRevB.42.8133} {\bibfield  {journal} {\bibinfo
  {journal} {Phys. Rev. B}\ }\textbf {\bibinfo {volume} {42}},\ \bibinfo
  {pages} {8133--8144} (\bibinfo {year} {1990})}\BibitemShut {NoStop}%
\bibitem [{\citenamefont {Yang}(2001)}]{Yang2001}%
  \BibitemOpen
  \bibfield  {author} {\bibinfo {author} {\bibfnamefont {K.}~\bibnamefont
  {Yang}},\ }\bibfield  {title} {\enquote {\bibinfo {title} {Dipolar excitons,
  spontaneous phase coherence, and superfluid-insulator transition in bilayer
  quantum hall systems at
  $\mathit{\ensuremath{\nu}}\phantom{\rule{0ex}{0ex}}=\phantom{\rule{0ex}{0ex}}1$},}\
  }\href {\doibase 10.1103/PhysRevLett.87.056802} {\bibfield  {journal}
  {\bibinfo  {journal} {Phys. Rev. Lett.}\ }\textbf {\bibinfo {volume} {87}},\
  \bibinfo {pages} {056802} (\bibinfo {year} {2001})}\BibitemShut {NoStop}%
\bibitem [{\citenamefont {Barkeshli}\ and\ \citenamefont
  {McGreevy}(2012)}]{Barkeshli2012}%
  \BibitemOpen
  \bibfield  {author} {\bibinfo {author} {\bibfnamefont {M.}~\bibnamefont
  {Barkeshli}}\ and\ \bibinfo {author} {\bibfnamefont {J.}~\bibnamefont
  {McGreevy}},\ }\bibfield  {title} {\enquote {\bibinfo {title} {Continuous
  transitions between composite fermi liquid and landau fermi liquid: A route
  to fractionalized mott insulators},}\ }\href {\doibase
  10.1103/PhysRevB.86.075136} {\bibfield  {journal} {\bibinfo  {journal} {Phys.
  Rev. B}\ }\textbf {\bibinfo {volume} {86}},\ \bibinfo {pages} {075136}
  (\bibinfo {year} {2012})}\BibitemShut {NoStop}%
\bibitem [{\citenamefont {Bargmann}(1962)}]{Bargmann1962}%
  \BibitemOpen
  \bibfield  {author} {\bibinfo {author} {\bibfnamefont {V.}~\bibnamefont
  {Bargmann}},\ }\bibfield  {title} {\enquote {\bibinfo {title} {On the
  representations of the rotation group},}\ }\href {\doibase
  10.1103/RevModPhys.34.829} {\bibfield  {journal} {\bibinfo  {journal} {Rev.
  Mod. Phys.}\ }\textbf {\bibinfo {volume} {34}},\ \bibinfo {pages} {829}
  (\bibinfo {year} {1962})}\BibitemShut {NoStop}%
\end{thebibliography}%

\clearpage
\onecolumngrid
\appendix

\renewcommand\thefigure{\thesection.\arabic{figure}}

\section{Sign of the pairing channel}
\setcounter{figure}{0}

TRS is explicitly broken in the CEL-CEL due to the external gauge field. 
Even though the CEs do not feel a net magnetic field anymore, the sign of the CS term in the HLR theory is dictated by the direction of the magnetic field. 
In the Eliashberg theory for the interlayer CEL-CEL pairing, this means that the $\pm l$ pairing channels are split and that which of the two is favored is determined by the sign of the CS term~\cite{Isobe2017}. 
At first thought, it might seem like there is no qualitative difference between the $\pm l$ pairings. 
However, because the direction of the magnetic field $B$ and the charge of the particles $q$ already fix the structure of the Jastrow factors, holomorphic coordinates for $qB>0$ and anti-holomorphic ones for $qB<0$, it does make a difference whether the pairing wave function has angular momentum symmetry $+l$ or $-l$. 
Here, we present arguments that should settle the confusion about the sign of the pairing channel.

The starting point is to use the correct Schr\"odinger Lagrangian for charged particles in an external gauge field. 
The corresponding equation of motion for the field of the charged particle $c$ should be
\begin{equation}
    i \partial_t c = \frac{1}{2m} \left(-i\vec\nabla - q\vec A\right)^2 c + q A_t c.
\end{equation}
We have set $\hbar = c = e = 1$, where $e$ is the elementary charge. The Lagrangian density is thus
\begin{equation}
    \mathcal{L} = \cd \left( i\partial_t - q A_t \right) c - \frac{1}{2m} \cd \left( i \vec\nabla + q \vec A \right)^2 c.
\end{equation}
Rewriting this in terms of the CF field $\psid = \cd e^{iq \int d\vec x' \arg(\vec x - \vec x') \rho(\vec x')}$ gives
\begin{equation}
    \mathcal{L} = \psid \left( i\partial_t - q A_t + q a_t\right) \psi - \frac{1}{2m} \psid \left( i \vec\nabla + q \vec A - q \vec a \right)^2 \psi - \frac{q}{4\pi} a_t \vec\nabla \wedge \vec a.
\end{equation}
Assume now that $q$ is positive. We know that the interlayer pairing symmetry in the case of a negative CS term is $l=+1$~\cite{Isobe2017}. A sensible BCS pairing wave function that has this symmetry is
\begin{equation}
    \det \left[ \left( \frac{1}{(\zup_i - \zdown_j)^{\ast}} \right)_{ij} \right].
\end{equation}
Here we have used holomorphic complex coordinates $z = x + iy$.
Playing the same game with negative $q$, i.e. the actual CEL-CEL case, the CS term is positive, and the pairing channel is $l=-1$. 
Importantly though, the wave function still has the same form since now we are using anti-holomorphic coordinates.

\section{PH duality}
\setcounter{figure}{0}

\subsection{PH transformation in the Dirac CF theory}

The Dirac CF theory proposed by Son~\cite{Son2015} is the fermionic particle-vortex dual theory~\cite{Seiberg2016} of Dirac electrons in a background field. 
Its Lagrangian density is given by
\begin{equation}
    \mathcal{L} = i \bar\psi \left( \fsl{\partial} + 2i\fsl{a} \right) \psi + \frac{1}{2\pi} Ada + \dots,
\end{equation}
where the $\dots$ stand for non-universal terms. The adopted convention for the $\gamma$-matrices is
\begin{equation}
    \gamma^0 = \sigma_z, \quad \gamma^1 = i\sigma_y, \quad \gamma^2 = -i\sigma_x.
\end{equation}
PH transformation in the Dirac CF theory amounts to a concatenation of charge conjugation $\mathcal{C}$ and time reversal $\mathcal{T}$, and is consequently anti-unitary. The transformation rules for the different fields are
\begin{align}
    \mathcal{CT} \psi(t,\vec x) (\mathcal{CT})^{-1} &= -i\sigma_y \psi(-t,\vec x),\\
    \mathcal{CT} a_0(t,\vec x) (\mathcal{CT})^{-1} &= a_0(-t,\vec x), \quad \mathcal{CT} a_i(t,\vec x) (\mathcal{CT})^{-1} = -a_i(-t,\vec x),\\
    \mathcal{CT} A_0(t,\vec x) (\mathcal{CT})^{-1} &= -A_0(-t,\vec x), \quad \mathcal{CT} A_i(t,\vec x) (\mathcal{CT})^{-1} = A_i(-t,\vec x).
\end{align}
Thus, the Lagrangian in the above form remains invariant under $\mathcal{CT}$. Moreover, a mass term $\bar\psi \psi$ or a pure CS term $ada$ change sign under the PH transformation and break this symmetry. While, a chemical potential term $\bar\psi \gamma^0 \psi$ remains invariant.

\subsection{CEL as non-relativistic limit}

We will quickly repeat how in Ref.~\cite{Son2015} the non-relativistic limit of the Dirac CF theory is taken. One starts by introducing a $\mathcal{CT}$-breaking mass term $-m \bar\psi \psi$ as well as a chemical potential $\mu \bar\psi \gamma^0 \psi$ to the Lagrangian. If the mass is large and the chemical potential lies within the gap, we can integrate out $\psi$ to obtain as an effective action for the gauge field
\begin{equation}
    -\frac{1}{2\pi} \frac{m}{|m|} ada.
\end{equation}
If the chemical potential now lies just above the gap, the gapless fermions are non-relativistic but still feel the effect of the filled Dirac sea. The appropriate theory then is
\begin{equation}
    \mathcal{L} = \psid (i\partial_t - 2 a_0) \psi - \frac{1}{2|m|} \psid (i\vec\nabla + 2\vec a)^2 \psi - \frac{1}{2\pi} \frac{m}{|m|} ada + \frac{1}{2\pi} Ada.
\end{equation}
Assuming a negative mass and making the substitution $2a \rightarrow a - A$, the resulting theory is
\begin{equation}
    \mathcal{L} = \psid (i\partial_t + A_0 - a_0) \psi - \frac{1}{2|m|} \psid (i\vec\nabla - \vec A + \vec a)^2 \psi + \frac{1}{8\pi} ada - \frac{1}{8\pi} AdA.
\end{equation}
This is the usual HLR formulation plus an additional CS term for the external field which is the contribution of the negative energy states to the Hall conductivity.
We should thus add a $\frac{1}{8\pi} AdA$ CS term to correct for this.

\subsection{CHL as non-relativistic limit}

As explained above, the $\mathcal{CT}$ transformation only changes the sign of the mass term.
Again making the approximation of having a chemical potential just above the mass gap then gives the effective theory
\begin{equation}
    \mathcal{L} = \psid (i\partial_t - 2 a_0) \psi - \frac{1}{2|m|} \psid (i\vec\nabla + 2\vec a)^2 \psi + \frac{1}{2\pi} \frac{m}{|m|} ada + \frac{1}{2\pi} Ada.
\end{equation}
Assuming a negative mass and making the substitution $2a \rightarrow A - a$ gives
\begin{equation}
    \mathcal{L} = \psid (i\partial_t - A_0 + a_0) \psi - \frac{1}{2|m|} \psid (i\vec\nabla + \vec A - \vec a)^2 \psi - \frac{1}{8\pi} ada + \frac{1}{8\pi} AdA.
\end{equation}
Adding the correction then leaves us with a total CS term of $\frac{1}{4\pi} AdA$ for the external gauge field.
This is the CS term accounting for the filled LLL in the hole theory~\cite{Barkeshli2012}.
The resulting theory is thus the CHL.

\subsection{Pairing and PH transformation}

The CEL-CHL is the non-relativistic limit of a massive bilayer Dirac CFL where one layer is PH-transformed.
We choose to apply the PH transformation in the bottom layer and the total transformation acts as
\begin{equation}
    \mathcal{CT}_{\text{b}} \psi (\mathcal{CT}_{\text{b}})^{-1} = \begin{pmatrix}
        \sigma_0 & 0\\
        0 & -i\sigma_y
    \end{pmatrix} \psi,
\end{equation}
where $\psi = \begin{pmatrix} \psit & \psib \end{pmatrix}^{\intercal}$.
This transforms the triplet-interlayer-pairing term as
\begin{equation}
    \mathcal{CT}_{\text{b}} \Delta \psi^{\intercal} i \sigma_y \tau_x \psi (\mathcal{CT}_{\text{b}})^{-1}
    = \Delta \psi^{\intercal} 
    \begin{pmatrix}
        \sigma_0 & 0\\
        0 & i\sigma_y
    \end{pmatrix}
    \begin{pmatrix}
        0 & i\sigma_y\\
        i\sigma_y & 0
    \end{pmatrix}
    \begin{pmatrix}
        \sigma_0 & 0\\
        0 & -i\sigma_y
    \end{pmatrix}
    \psi
    = \Delta \psi^{\intercal} \sigma_0 i\tau_y \psi.
\end{equation}
Similarly, for the singlet-interlayer-pairing term we have
\begin{equation}
    \mathcal{CT}_{\text{b}} \Delta \psi^{\intercal} i\sigma_y i\tau_y \psi (\mathcal{CT}_{\text{b}})^{-1}
    = \Delta \psi^{\intercal} 
    \begin{pmatrix}
        \sigma_0 & 0\\
        0 & i\sigma_y
    \end{pmatrix}
    \begin{pmatrix}
        0 & i\sigma_y\\
        -i\sigma_y & 0
    \end{pmatrix}
    \begin{pmatrix}
        \sigma_0 & 0\\
        0 & -i\sigma_y
    \end{pmatrix}
    \psi
    = \Delta \psi^{\intercal} \sigma_0 \tau_x \psi,
\end{equation}
where now to ensure the total fermionic anti-symmetry we assumed the gap function to be odd in the momentum $\vec k$.

\subsection{BdG calculations}

To obtain the correct pairing terms in the non-relativistic theories, we need to project the pairings to the respective Fermi surfaces. We start with the CEL-CEL case. The full pairing Hamiltonian in the Dirac-CF theory is
\begin{equation}
    \mathcal{H} = \sum_{s=\text{t,b}} \psid_s \, (\vec p \cdot \vec \sigma - \mu) \, \psi_s + \left( i\Delta \psit \sigma_y \psib + i\Delta \psib \sigma_y \psit + \text{h.c.} \right),
\end{equation}
where $s=\text{t,b}$ denotes the top and bottom layer, and we use $\uparrow / \downarrow$ for the internal spinor structure of the Dirac fields. The corresponding BdG Hamiltonian in the basis $(\psit(\vec k), \psib(\vec k), \psitd(-\vec k), \psibd(-\vec k))^{\intercal}$ is
\begin{equation}
    \mathcal{H}_{\text{BdG}} = 
    \begin{pmatrix}
        \vec k \cdot \vec\sigma- \mu & 0 & 0 & -i\Delta \sigma_y\\
        0 & \vec k \cdot \vec\sigma- \mu & -i\Delta \sigma_y & 0\\
        0 & i\Delta \sigma_y & \vec k \cdot \vec\sigma^{\ast} + \mu & 0\\
        i\Delta \sigma_y & 0 & 0 & \vec k \cdot \vec\sigma^{\ast} + \mu
    \end{pmatrix}.
\end{equation}
The interlayer pairing has triplet structure $\tau_x$ and the internal pseudospin has singlet structure $i\sigma_y$. We can isolate the interlayer pairing elements
\begin{equation}
    \mathcal{H}_{\text{BdG}}^{\text{tb}} =
    \begin{pmatrix}
        \vec k \cdot \vec\sigma- \mu & -i\Delta \sigma_y\\
        i\Delta \sigma_y & \vec k \cdot \vec\sigma^{\ast} + \mu
    \end{pmatrix}.
\end{equation}
The diagonal part has a free Dirac dispersion $E_{\pm}(\vec k) = \pm|\vec k| - \mu$ and eigenstates $\psi_{\pm}(\vec k) = 1/\sqrt{2} \left( \psiup(\vec k) \pm e^{i\theta(\vec k)} \psidown(\vec k) \right)$, where $\theta(\vec k)$ is the polar angle of $\vec k$. To transform the pairing part into this new basis we write
\begin{align}
\begin{split}
    & \begin{pmatrix} \psitupd(\vec k) & \psitdownd(\vec k) \end{pmatrix} i\Delta \sigma_y \begin{pmatrix} \psibupd(-\vec k)\\ \psibdownd(-\vec k) \end{pmatrix}\\
    =& \begin{pmatrix} \psid_{\text{t}+}(\vec k) & \psid_{\text{t}-}(\vec k) \end{pmatrix}
    \frac{1}{\sqrt{2}}\begin{pmatrix}
        1 & e^{i\theta(\vec k)}\\
        1 & -e^{i\theta(\vec k)}
    \end{pmatrix} i\Delta \sigma_y \frac{1}{\sqrt{2}} \begin{pmatrix}
        1 & 1\\
        e^{i\theta(-\vec k)} & -e^{i\theta(-\vec k)}
    \end{pmatrix}
    \begin{pmatrix} \psid_{\text{b}+}(-\vec k)\\ \psid_{\text{b}-}(-\vec k) \end{pmatrix}\\
    =& \begin{pmatrix} \psid_{\text{t}+}(\vec k) & \psid_{\text{t}-}(\vec k) \end{pmatrix}
    \frac{1}{\sqrt{2}}\begin{pmatrix}
        1 & e^{i\theta(\vec k)}\\
        1 & -e^{i\theta(\vec k)}
    \end{pmatrix} \Delta \frac{1}{\sqrt{2}} \begin{pmatrix}
        -e^{i\theta(\vec k)} & e^{i\theta(\vec k)}\\
        -1 & -1
    \end{pmatrix}
    \begin{pmatrix} \psid_{\text{b}+}(-\vec k)\\ \psid_{\text{b}-}(-\vec k) \end{pmatrix}\\
    =& \begin{pmatrix} \psid_{\text{t}+}(\vec k) & \psid_{\text{t}-}(\vec k) \end{pmatrix}
    \Delta \begin{pmatrix}
        -e^{i\theta(\vec k)} & 0\\
        0 & e^{i\theta(\vec k)}
    \end{pmatrix}
    \begin{pmatrix} \psid_{\text{b}+}(-\vec k)\\ \psid_{\text{b}-}(-\vec k) \end{pmatrix}.
\end{split}
\end{align}
In the limit $\mu \gg |\Delta|$, we only retain the conduction band, and the relevant pairing term $\Delta e^{i\theta(\vec k)} \, \psid_{\text{t}+}(\vec k) \psid_{\text{b}+}(-\vec k)$ leads to a $p_x + ip_y$ pairing structure, similar to the celebrated Fu-Kane mechanism of chiral topological superconductivity on the surface of a topological insulator~\cite{Fu2008}. 
The only differences are: (i) our Dirac fields $\psi$ correspond to composite fermions, and (ii) we have double degeneracy due to the second identical copy.

To similarly obtain the band-projected pairing in the CEL-CHL representation, we start with the full pairing Hamiltonian, when the Dirac CF of the bottom layer is transformed under a $\mathcal{CT}$ transformation. 
\begin{equation}
    \mathcal{H} = \sum_{s=\text{t,b}} \psid_s \, (\vec p \cdot \vec \sigma - \mu) \, \psi_s + \left( \Delta \psit \sigma_0 \psib - \Delta \psib \sigma_0 \psit + \text{h.c.} \right),
\end{equation}
and the BdG Hamiltonian is
\begin{equation}
    \mathcal{H}_{\text{BdG}} = 
    \begin{pmatrix}
        \vec k \cdot \vec\sigma- \mu & 0 & 0 & -\Delta \sigma_0\\
        0 & \vec k \cdot \vec\sigma- \mu & \Delta \sigma_0 & 0\\
        0 & \Delta \sigma_0 & \vec k \cdot \vec\sigma^{\ast} + \mu & 0\\
        -\Delta \sigma_0 & 0 & 0 & \vec k \cdot \vec\sigma^{\ast} + \mu
    \end{pmatrix}.
\end{equation}
Opposite to the previous case, this representation leads to the pairing having interlayer singlet structure $i\tau_y$ and internal pseudospin triplet structure $\sigma_0$. The interlayer pairing elements are
\begin{equation}
    \mathcal{H}_{\text{BdG}}^{\text{tb}} =
    \begin{pmatrix}
        \vec k \cdot \vec\sigma- \mu & -\Delta \sigma_0\\
        -\Delta \sigma_0 & \vec k \cdot \vec\sigma^{\ast} + \mu
    \end{pmatrix}.
\end{equation}
The single-particle eigenstates are the same as before, only that we need to keep in mind that $\psib$ is evaluated at negative time compared with $\psit$ and thus we need to apply a complex conjugation $\mathcal{K}$. 
Hence, the Dirac eigenstates in the bottom layer are now $\psi_{\pm}(\vec k) = 1/\sqrt{2} \left( \psiup(\vec k) \pm e^{-i\theta(\vec k)} \psidown(\vec k) \right)$. We can again transform the pairing into this basis
\begin{align}
\begin{split}
    & \begin{pmatrix} \psitupd(\vec k) & \psitdownd(\vec k) \end{pmatrix} \Delta \sigma_0 \begin{pmatrix} \psibupd(-\vec k)\\ \psibdownd(-\vec k) \end{pmatrix}\\
    =& \begin{pmatrix} \psid_{\text{t}+}(\vec k) & \psid_{\text{t}-}(\vec k) \end{pmatrix}
    \frac{1}{\sqrt{2}}\begin{pmatrix}
        1 & e^{i\theta(\vec k)}\\
        1 & -e^{i\theta(\vec k)}
    \end{pmatrix} \Delta \sigma_0 \frac{1}{\sqrt{2}} \begin{pmatrix}
        1 & 1\\
        -e^{-i\theta(\vec k)} & e^{-i\theta(\vec k)}
    \end{pmatrix}
    \begin{pmatrix} \psid_{\text{b}+}(-\vec k)\\ \psid_{\text{b}-}(-\vec k) \end{pmatrix}\\
    =& \begin{pmatrix} \psid_{\text{t}+}(\vec k) & \psid_{\text{t}-}(\vec k) \end{pmatrix}
    \Delta \begin{pmatrix}
        0 & 1\\
        1 & 0
    \end{pmatrix}
    \begin{pmatrix} \psid_{\text{b}+}(-\vec k)\\ \psid_{\text{b}-}(-\vec k) \end{pmatrix}.
\end{split}
\end{align}
Projection onto the Fermi surface just as above leaves us with an $s$-wave superconductor; since we inverted the bottom layer Dirac cone, $\psi_{\text{b}-}$ is now at the Fermi surface.

\section{Wave functions}
\setcounter{figure}{0}

\subsection{PH transformation in the LLL}

For a wave function $\psi_{\text{particles}}(z_1,\dots,z_N)$ of $N$ particles in the LLL with a total degeneracy of $M>N$, its PH-conjugate wave function $\Theta[\psi_{\text{particles}}]$ in terms of $M-N$ holes is~\cite{Girvin1984,Yang2001}
\begin{equation}
    \psi_{\text{holes}}(w_1,\dots,w_{M-N}) = \int \prod_i^N dz_i dz_i^{\ast} \left[ \prod_{i<j}^N (z_i-z_j)^{\ast} \prod_{i<j}^{M-N} (w_i-w_j)^{\ast} \prod_i^N \prod_j^{M-N} (z_i-w_j)^{\ast} \right] \psi_{\text{particles}}(z_1,\dots,z_N),
\end{equation}
where the integration measure is defined as $dzdz^{\ast} = e^{-|z|^2 / 2\ell^2} \frac{dxdy}{2\pi \ell^2}$. It is important to realize that the coordinates of the holes only appear in the form $w_i^{\ast}$. This is exactly what one should expect because the holes and the particles are oppositely charged and thus their Aharonov-Bohm phases in the external field are opposite as well, which manifests itself in having holomorphic coordinates for one and anti-holomorphic coordinates for the other. The inverse transform, i.e. starting from the hole wave function $\psi_{\text{holes}}(w_1,\dots,w_{M-N})$, is
\begin{equation}
    \psi_{\text{particles}}(z_1,\dots,z_N) = \int \prod_i^{M-N} dw_i dw_i^{\ast} \left[ \prod_{i<j}^N (z_i-z_j) \prod_{i<j}^{M-N} (w_i-w_j) \prod_i^N \prod_j^{M-N} (z_i-w_j) \right] \psi_{\text{holes}}(w_1,\dots,w_{M-N}).
\end{equation}

\subsection{Exciton condensate at $d = 0$}

\begin{figure}[h]
    \centering
    \includegraphics[height=0.3\textwidth]{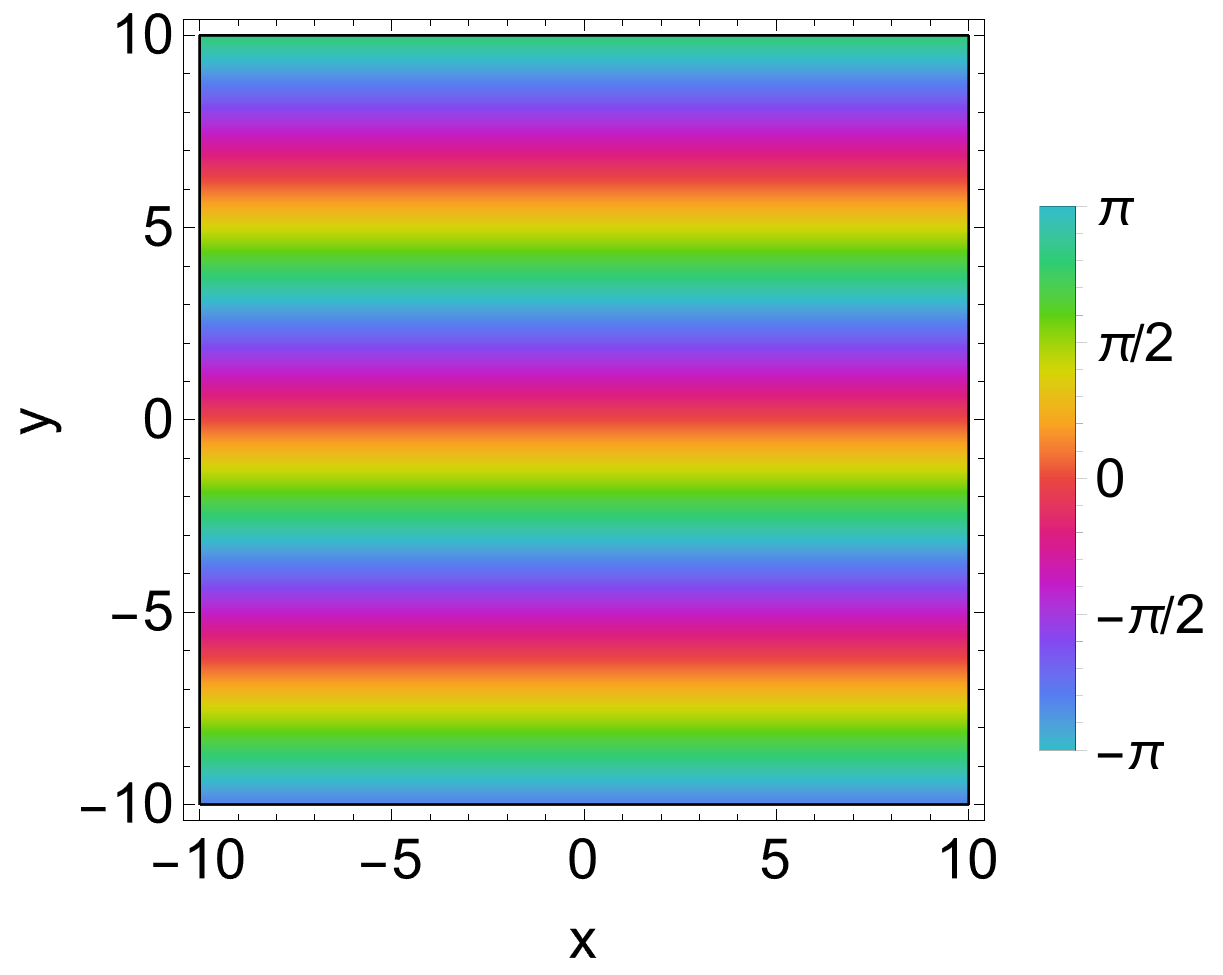}
    \caption{Phase structure of $e^{\zup (\wdown)^{\ast} / 2\ell^2}$ with fixed $\wdown = -1$ and $\zup$ as free variable}
    \label{figapp: phase plots pairing xc}
\end{figure}

We want to derive an electron-hole wave function for the QHB at $d=0$. The electron-electron wave function describing this phase is the Halperin~(111) wave function~\cite{Halperin1983}
\begin{equation}
\label{eqapp: halperin 111}
    \psi_{111} = \prod_{i<j} (\zup_i - \zup_j) \prod_{i<j} (\zdown_i - \zdown_j) \prod_{ij} (\zup_i - \zdown_j).
\end{equation}
Instead of directly trying to transform this wave function, we take the commonly used XC wave function~\cite{Yang2001}
\begin{equation}
    \psi_{\text{XC}} = \det \left[ \left( e^{\zup_i (\wdown_j)^{\ast} / 2\ell^2} \right)_{ij} \right],
\end{equation}
and transform the holes in the bottom layer $\wdown$ back into particles $\zdown$. We obtain
\begin{equation}
    \int \prod_i^{N} d\wdown_i d(\wdown_i)^{\ast} \left[ \prod_{i<j}^N (\zdown_i-\zdown_j) \prod_{i<j}^{N} (\wdown_i-\wdown_j) \prod_i^N \prod_j^{N} (\zdown_i-\wdown_j) \right] \psi_{\text{XC}} = \prod_{i<j} (\zup_i - \zup_j) \prod_{i<j} (\zdown_i - \zdown_j) \prod_{ij} (\zup_i - \zdown_j),
\end{equation}
where because we are at half-filling $M=2N$, and we have made use of the Bargmann identity~\cite{Bargmann1962} for holomorphic functions $f(z)$
\begin{equation}
    \int dwdw^{\ast} e^{zw^{\ast} / 2\ell^2} f(w) = f(z).
\end{equation}
This shows that PH transforming one layer in $\psi_{111}$ leads to $\psi_{\text{XC}}$. To see that $\psi_{\text{XC}}$ indeed describes electron-hole pairs, we multiply the modulus squared of one of the entries in the determinant with its corresponding exponential factors
\begin{equation}
    \left|e^{\zup_i (\wdown_j)^{\ast} / 2\ell^2}\right|^2 e^{-|z_i|^2/4\ell^2} e^{-|w_j|^2/4\ell^2} = e^{-|z_i-w_j|^2/4\ell^2}.
\end{equation}
To see that $\psi_{\text{XC}}$ has $s$-wave pairing symmetry we look at the two-particle pairing function $e^{\zup (\wdown)^{\ast} / 2\ell^2}$.
In Fig.~\ref{figapp: phase plots pairing xc}, we plot the phase structure of this pairing function where we fixed the hole as $\wdown = -1$ and the electron $\zup$ is the free variable.
Clearly, as we take the electron around the position of the hole, no phase is picked up.

\subsection{CF phase at large-$d$}

Coming from the $d \rightarrow \infty$ limit where the two layers are entirely decoupled, they are described by CFLs. The bilayer CFL has weak-coupling BCS instabilities: in the $p$-wave channel for a CEL-CEL description~\cite{Isobe2017}, and in the $s$-wave channel for a CEL-CHL~\cite{Rueegg2023}. 
A wave function that captures the CE-CE pairing with $l = 1$ angular-momentum symmetry can be written as~\cite{Sodemann2017}:
\begin{equation}
    \psi_{\text{CEL-CEL,p}} = \prod_{i < j}(\zup_i - \zup_j)^2 \prod_{i < j}(\zdown_i - \zdown_j)^2 \, \det \left[ \left( \frac{1}{(\zup_i - \zdown_j)^{\ast}} \right)_{ij} \right].
\end{equation}
The Jastrow factors in front capture the CF nature and the determinant incorporates the interlayer two-particle pairing function. Using the determinant of the Cauchy matrix
\begin{equation}
    \det \left[ \left( \frac{1}{(\zup_i - \zdown_j)^{\ast}} \right)_{ij} \right] = \prod_{i,j} \frac{1}{(\zup_i - \zdown_j)^{\ast}} \prod_{i<j} (\zup_i - \zup)^{\ast} \prod_{i<j} (\zdown_i - \zdown_j)^{\ast},
\end{equation}
the wave function can be rewritten as
\begin{equation}
    \psi_{\text{CEL-CEL,p}} = \prod_{i,j} \frac{1}{|\zup_i - \zdown_j|^2} \prod_{i<j} |\zup_i - \zup|^2 \prod_{i<j} |\zdown_i - \zdown_j|^2 \, \psi_{111}.
\end{equation}
From this, as well as field theory calculations, it can be argued that the large-$d$ phase is topologically connected with the small-$d$ XC. Indeed, Sodemann et al.~\cite{Sodemann2017} hint at the possibility of a BCS-BEC crossover.

In the same spirit, we consider a wave function for the CE-CH pairing with $l=0$ angular-momentum symmetry
\begin{equation}
    \psi_{\text{CEL-CHL,s}} = \prod_{i<j} (\zup_i - \zup_j)^2 \prod_{i<j} (\wdown_i - \wdown_j)^{\ast 2} \det \left[ \left( \prod_{k \neq i}(\zup_i - \zup_k)^{\ast} \prod_{k \neq j}(\wdown_j - \wdown_k) \, e^{\zup_i (\wdown_j)^{\ast} / 2\ell^2} \right)_{ij} \right].
\end{equation}
We realize that from the $i$th row in the determinant we can extract the factor $\prod_{k\neq i} (\zup_i - \zup_k)^{\ast}$, and analogously for $\prod_{k\neq j} (\wdown_j - \wdown_k)$ in the $j$th column. Since each Jastrow factor $(\zup_i - \zup_j)^{\ast}$ features in rows $i$ and $j$, we can write
\begin{equation}
    \psi_{\text{CEL-CHL,s}} = \prod_{i<j} |\zup_i - \zup_j|^4 \prod_{i<j} |\wdown_i - \wdown_j|^4 \, \psi_{\text{XC}}.
\end{equation}
Hence, we obtain wave functions with the same phase structure for the large- and the small-$d$ limit.

\begin{figure}[h]
    \centering
    \includegraphics[height=0.3\textwidth]{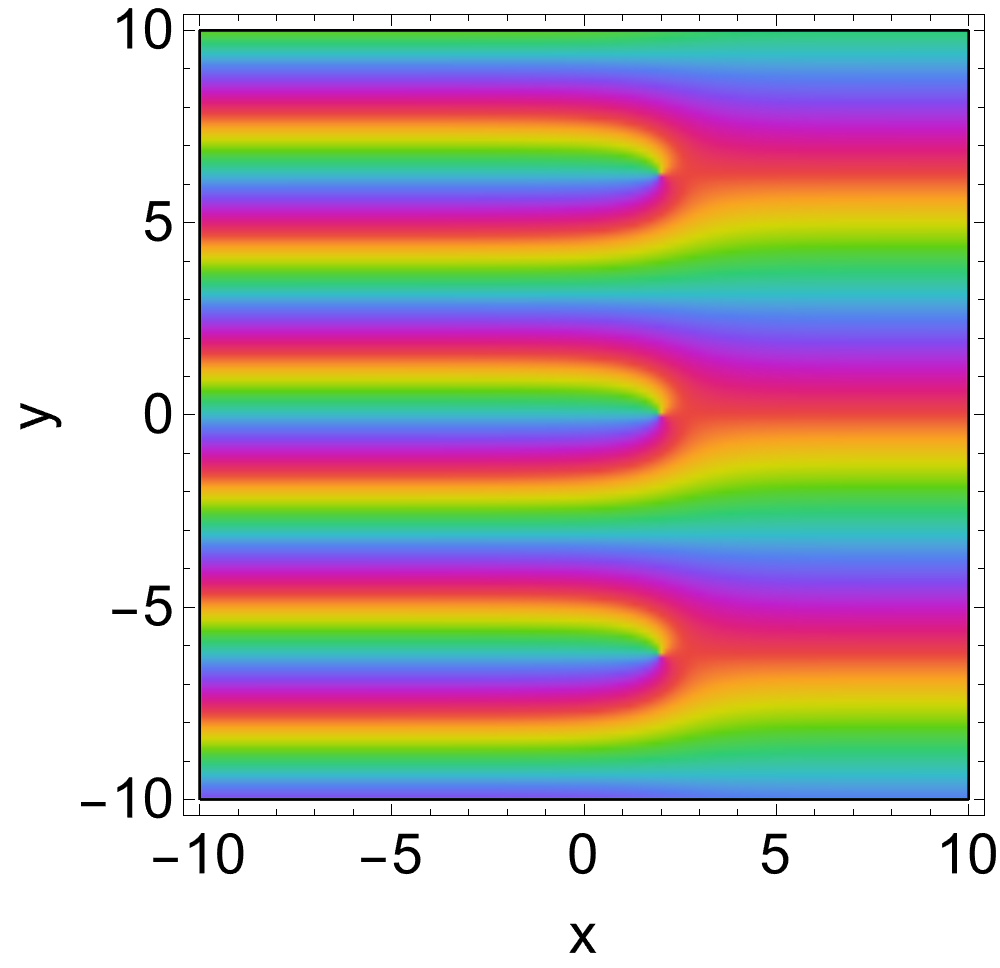}
    \quad
    \includegraphics[height=0.3\textwidth]{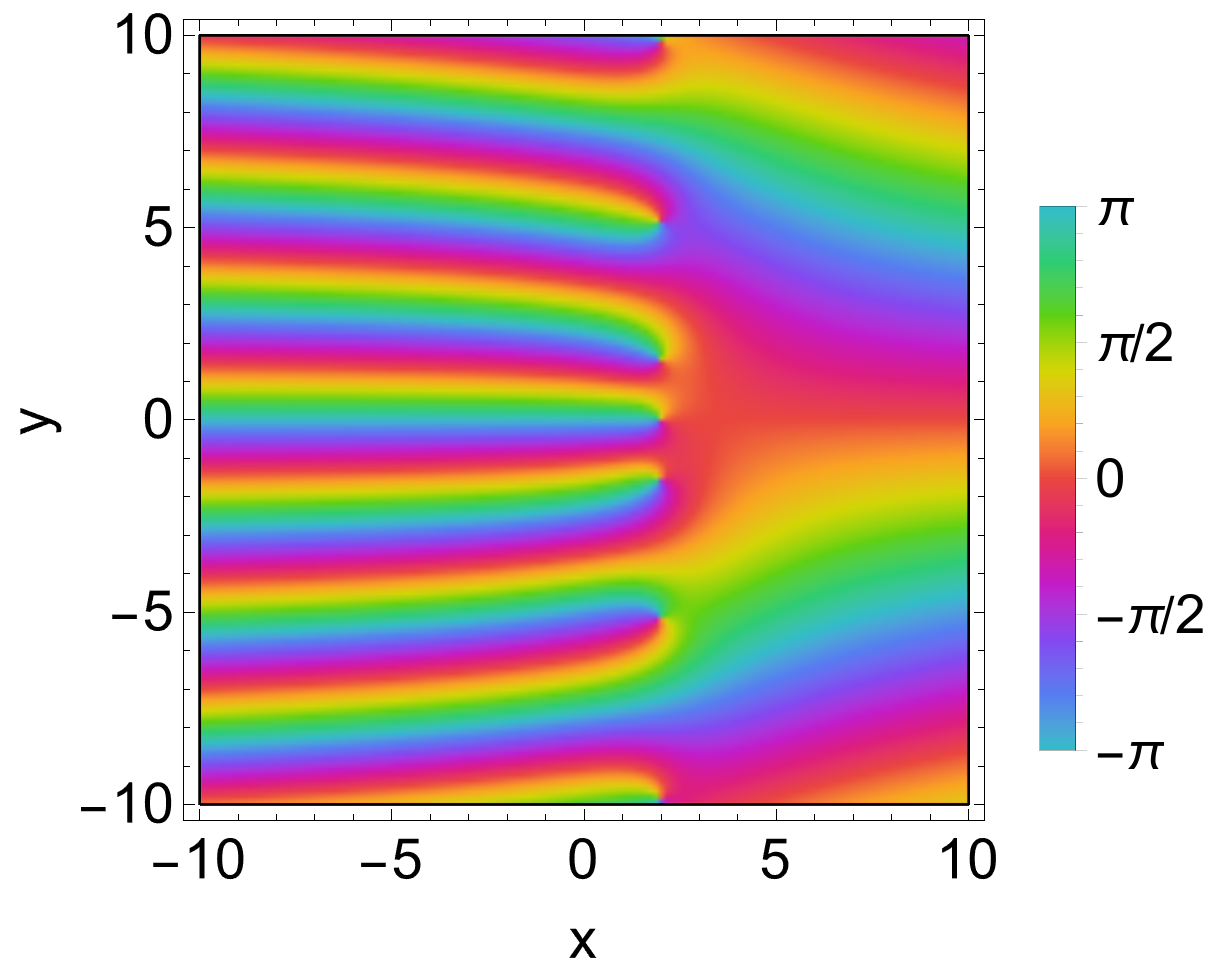}
    \caption{Phase structures of (left) $\psi_{\text{111}} (\zup_1,\zup_2=2,\wdown_1=-1,\wdown_2=-2)$ and (right) $\text{P}_{\text{LLL}} \psi_{\text{CEL-CHL,s}} (\zup_1,\zup_2=2,\wdown_1=-1,\wdown_2=-2)$}
    \label{figapp: phase plots xc and plllcelchls}
\end{figure}

To derive a practical trial wave function from it, we first need to project it into the LLL. This is done by setting $z^{\ast} \rightarrow \partial_z, \, w \rightarrow \partial_{w^{\ast}}$ and then taking all the partial derivatives to act on the rest of the wave function. We get
\begin{equation}
    \text{P}_{\text{LLL}} \psi_{\text{CEL-CHL,s}} = \prod_{i<j} (\partial_{\zup_i} - \partial_{\zup_j})^2 \prod_{i<j} (\partial_{(\wdown_i)^\ast} - \partial_{(\wdown_j)^\ast})^2 \prod_{i<j} (\zup_i - \zup_j)^2 \prod_{i<j} (\wdown_i - \wdown_j)^{\ast 2} \det \left[ \left( e^{\zup_i (\wdown_j)^{\ast} / 2\ell^2} \right)_{ij} \right].
\end{equation}
We evaluate this wave function for two particles in the top layer and two holes in the bottom layer. The comparison of the resulting phase structure with the one of $\psi_{111}$ is in Fig.~\ref{figapp: phase plots xc and plllcelchls}. There are two additional zeros close to the fermionic intralayer zero on the real axis. The introduction of new zeros in LLL-projected wave functions is a general feature.

\subsection{PH transformation of $\psi_{\text{CEL-CHL,s}}$}

Next, we show our proposed wave function's structure in the electron-electron basis. For this, we apply a PH transformation in the bottom layer.
\begin{align}
\begin{split}
    & (\Theta_{\text{PH}}^{\downarrow})^{-1} \left[\text{P}_{\text{LLL}} \psi_{\text{CEL-CHL,s}}\right]\\
    = &\int \prod_i^N d\wdown_i d(\wdown_i)^{\ast} \, \text{P}_{\text{LLL}} \psi_{\text{CEL-CHL,s}} (\zup_1,\dots,\zup_N,\wdown_1,\dots,\wdown_N) \, \psi_{\text{filled LLL}}(\wdown_1,\dots,\wdown_N,\zdown_1,\dots,\zdown_N)\\
    = & \prod_{i<j} (\partial_{\zup_i}-\partial_{\zup_j})^2 \prod_{i<j} (\zup_i-\zup_j)^2 \, \int \prod_i^N d\wdown_i d(\wdown_i)^{\ast} \, \prod_{i<j}(\partial_{(\wdown_i)^{\ast}}-\partial_{(\wdown_j)^{\ast}})^2 (\wdown_i-\wdown_j)^{\ast 2}\\
    & \sum_{\sigma} \sign(\sigma) \prod_i e^{\zup_{\sigma_i} (\wdown_i)^{\ast} / 2\ell^2} \prod_{i<j} (\wdown_i-\wdown_j) \prod_{i<j} (\zdown_i-\zdown_j) \prod_{i,j} (\zdown_i-\wdown_j)\\
    = & \prod_{i<j} (\partial_{\zup_i}-\partial_{\zup_j})^2 \prod_{i<j} (\zup_i-\zup_j)^2 \prod_{i<j} (\partial_{\zup_i}-\partial_{\zup_j})^2 \prod_{i<j} (\zup_i-\zup_j)^3 \prod_{i<j} (\zdown_i-\zdown_j) \prod_{i,j} (\zdown_i-\zup_j),
\end{split}
\end{align}
where we have eliminated the sum over all permutations by accordingly permuting all the $\wdown_i$ and since the total function is anti-symmetric this also eliminates the $\sign(\sigma)$, and we have again used the Bargmann identity for holomorphic functions. The commuting properties of the LLL projection and PH transformation can be checked by operating them in the opposite order to the above and obtain the same result:
\begin{align}
\begin{split}
    & \text{P}_{\text{LLL}} (\Theta_{\text{PH}}^{\downarrow})^{-1} \left[\psi_{\text{CEL-CHL,s}}\right]\\
    = &\text{P}_{\text{LLL}} \int \prod_i^N d\wdown_i d(\wdown_i)^{\ast} \, \psi_{\text{CEL-CHL,s}} (\zup_1,\dots,\zup_N,\wdown_1,\dots,\wdown_N) \, \psi_{\text{filled LLL}}(\wdown_1,\dots,\wdown_N,\zdown_1,\dots,\zdown_N)\\
    = & \text{P}_{\text{LLL}} \prod_{i<j} |\zup_i-\zup_j|^4 \, \int \prod_i^N d\wdown_i d(\wdown_i)^{\ast} \, \prod_{i<j}(\wdown_i-\wdown_j)^2 (\wdown_i-\wdown_j)^{\ast 2}\\
    & \sum_{\sigma} \sign(\sigma) \prod_i e^{\zup_{\sigma_i} (\wdown_i)^{\ast} / 2\ell^2} \prod_{i<j} (\wdown_i-\wdown_j) \prod_{i<j} (\zdown_i-\zdown_j) \prod_{i,j} (\zdown_i-\wdown_j)\\
    = & \prod_{i<j} (\partial_{\zup_i}-\partial_{\zup_j})^2 \prod_{i<j} (\zup_i-\zup_j)^2 \prod_{i<j} (\partial_{\zup_i}-\partial_{\zup_j})^2 \prod_{i<j} (\zup_i-\zup_j)^3 \prod_{i<j} (\zdown_i-\zdown_j) \prod_{i,j} (\zdown_i-\zup_j).
\end{split}
\end{align}
The resulting wave function is not layer-symmetric.
This can easily be resolved by taking a layer-symmetric combination
\begin{equation}
\label{eqapp: layer-symmetric PH-transformation}
    (\Theta^{\downarrow})^{-1}[\psi_{\text{CEL-CHL,s}}] + (\Theta^{\uparrow})^{-1}[\psi_{\text{CHL-CEL,s}}].
\end{equation}
Such a layer-symmetrization procedure has also been used in Ref.~\cite{Lian2018}.

In Fig.~\ref{figapp: phase plots plllphcelchls, quartic 111 and pllvarcelcelp}, we plot the phases of the LLL-projected PH transformed CEL-CHL $s$-wave on the left, and we compare it with $\text{P}_{\text{LLL}} \prod_{i<j} |\zup_i-\zup_j|^4 \, \psi_{111}$ in the middle and $\text{P}_{\text{LLL}} \prod_{i<j} |\zup_i - \zup_j|^2 \prod_{i<j} |\zdown_i - \zdown_j|^2 \prod_{i,j} |\zup_i - \zdown_j|^2 \psi_{\text{CEL-CEL,p}}$ on the right; the latter corresponds to the $n=2, m=4$ variational factor in Ref.~\cite{Sodemann2017}. All wave functions have the fermionic intralayer zeros, as well as something close to the interlayer zeros of the Halperin~(111) wave function. They also have the identical number of characteristic additional zeros of LLL-projected wave functions.
\begin{figure}[h]
    \centering
    \includegraphics[height=0.13\textwidth]{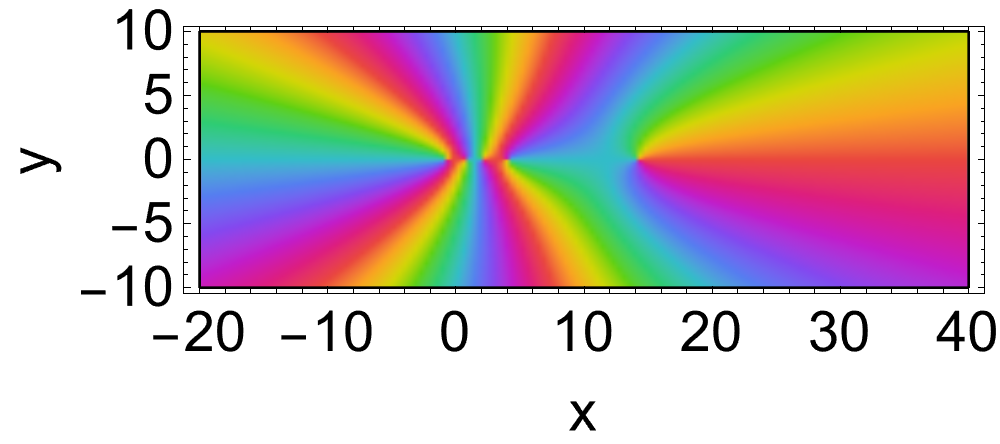}
    \quad
    \includegraphics[height=0.13\textwidth]{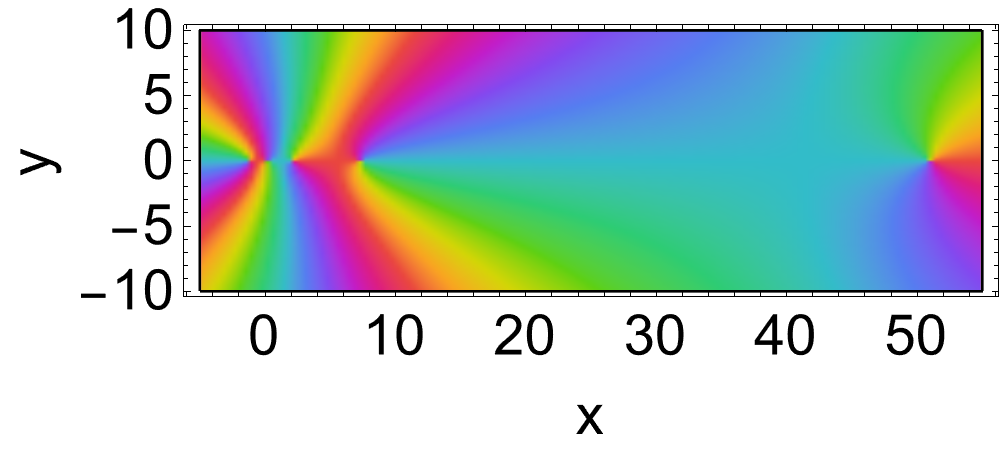}
    \quad
    \includegraphics[height=0.13\textwidth]{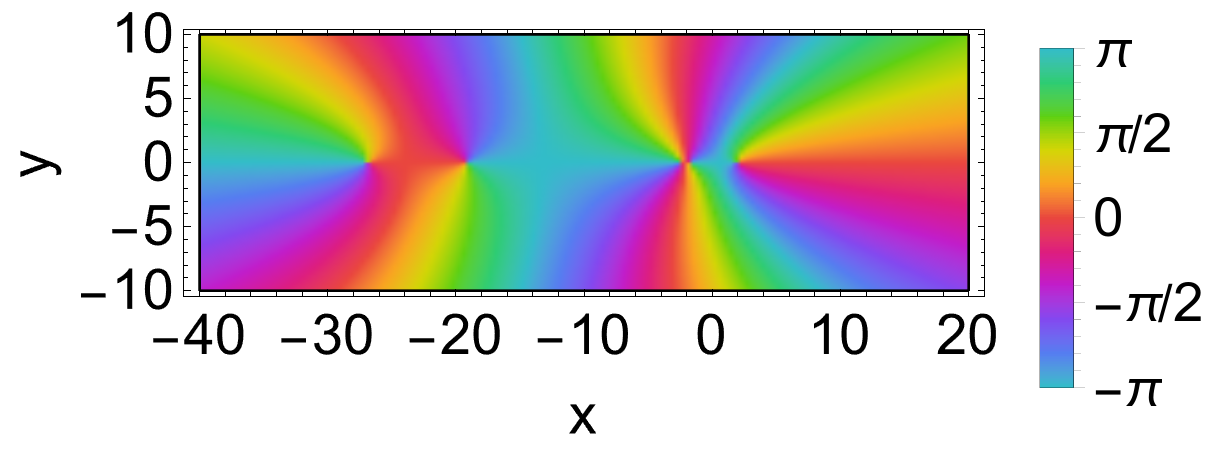}
    \caption{Phase structures of (left) $\text{P}_{\text{LLL}} (\Theta_{\text{PH}}^{\downarrow})^{-1} \left[\psi_{\text{CEL-CHL,s}}\right] (\zup_1,\zup_2=2,\wdown_1=-1,\wdown_2=-2)$, (middle) $\text{P}_{\text{LLL}} \prod_{i<j} |\zup_i-\zup_j|^4 \, \psi_{\text{111}} (\zup_1,\zup_2=2,\wdown_1=-1,\wdown_2=-2)$ and (right) $\text{P}_{\text{LLL}} \prod_{i<j} |\zup_i - \zup_j|^2 \prod_{i<j} |\zdown_i - \zdown_j|^2 \prod_{i,j} |\zup_i - \zdown_j|^2 \psi_{\text{CEL-CEL,p}}$}
    \label{figapp: phase plots plllphcelchls, quartic 111 and pllvarcelcelp}
\end{figure}

\subsection{CB phase at intermediate $d$}

Ref.~\cite{Lian2018} proposed an intermediate CB XC phase. Due to the reduced intralayer effective charge of the CBs the interlayer CB excitons can become charge-neutral around $d\approx l$. Thus, these CB electron-hole bound states can then condense. The wave function should be the product of the CB Jastrow factors and the bosonic condensate part. In Ref.~\cite{Lian2018}, it was argued that the natural form of the condensate wave function would be a permanent, the totally symmetric version of the determinant. Thus, 
\begin{equation}
    \psi_{\text{CB XC}}^{(1)} = \prod_{i<j} (\zup_i - \zup_j) \prod_{i<j} (\wdown_i - \wdown_j)^{\ast} \text{perm} \left[ \left( e^{\zup_i (\wdown_j)^{\ast} / 4\ell^2} \right)_{ij} \right].
\end{equation}
This wave function also shows peaked overlap with ED calculations at $d\approx l$.

Motivated by our method of absorbing conjugated Jastrow factors into the pairing function, we propose another wave function for the intermediate CB XC phase
\begin{equation}
    \psi_{\text{CB XC}}^{(2)} = \prod_{i<j} (\zup_i - \zup_j) \prod_{i<j} (\wdown_i - \wdown_j)^{\ast} \, \det \left[ \left( \prod_{k > i}(\zup_i - \zup_k)^{\ast} \prod_{k > j}(\wdown_j - \wdown_k) \, e^{\zup_i (\wdown_j)^{\ast} / 4\ell^2} \right)_{ij} \right].
\end{equation}
Every Jastrow factor $(\zup_i - \zup_j)$ / $(\wdown_i - \wdown_j)^{\ast}$ now only appears in one row/column which is why there is no double-counting this time. Thus, the determinant is a bosonic pairing function. This wave function again has the same phase structure as $\psi_{\text{XC}}$. In Fig.~\ref{figapp: phase plots xc2, cbxcbiao and cbxc2} we compare the phase structures of $\psi_{\text{XC}}$ on the left, $\psi_{\text{CB XC}}^{(1)}$ in the middle, and $\text{P}_{\text{LLL}} \, \psi_{\text{CB XC}}^{(2)}$ on the right. It is interesting to see that the zeros in all three wave functions are at similar positions, and that the two CB wave functions have almost identical phase structures.
\begin{figure}[h]
    \centering
    \includegraphics[height=0.25\textwidth]{images/plotXC2.png}
    \quad
    \includegraphics[height=0.25\textwidth]{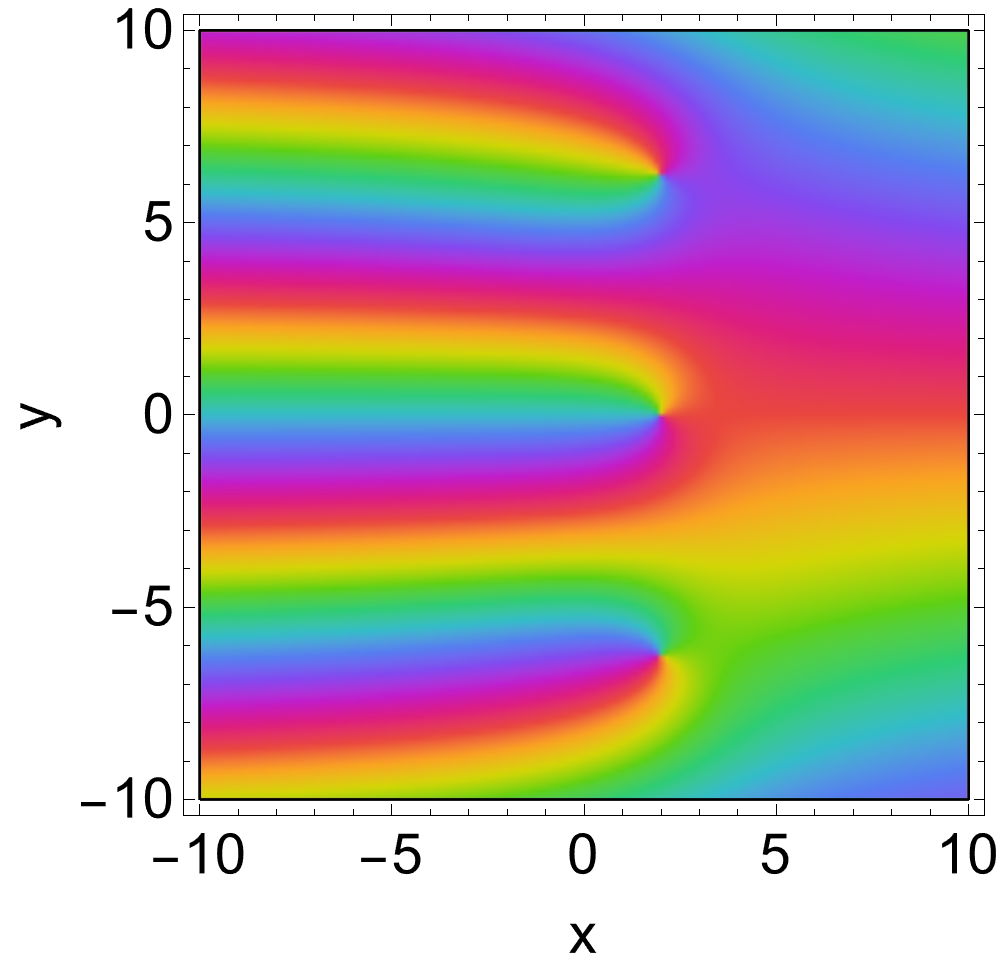}
    \quad
    \includegraphics[height=0.25\textwidth]{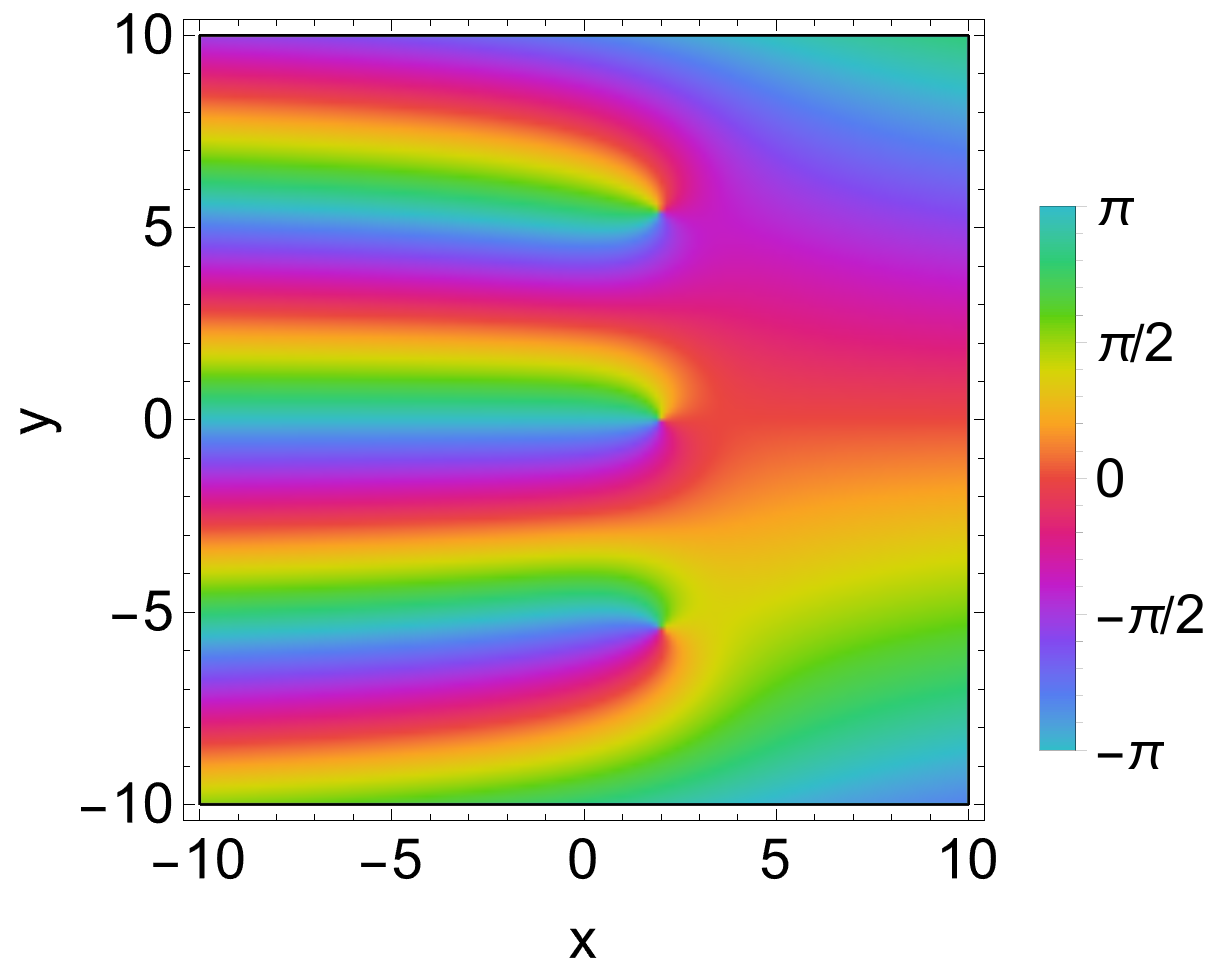}
    \caption{Phases of the wave functions (left) $\psi_{\text{XC}}(\zup_1, \zup_2 = 2, \zdown_1 = -1, \zdown_2 = -2)$, (middle) $\psi_{\text{CB XC}}^{(1)}(\zup_1, \zup_2 = 2, \zdown_1 = -1, \zdown_2 = -2)$ and (right) $\text{P}_{\text{LLL}} \, \psi_{\text{CB XC}}^{(2)}(\zup_1, \zup_2 = 2, \zdown_1 = -1, \zdown_2 = -2)$}
    \label{figapp: phase plots xc2, cbxcbiao and cbxc2}
\end{figure}

\subsection{PH transformation of $\psi_{\text{CB XC}}^{(2)}$}

Next, we apply PH transformation in the bottom layer to see the form of $\psi^{(2)}_{\text{CB XC}}$ in the electron-electron basis.
\begin{align}
\begin{split}
    & (\Theta_{\text{PH}}^{\downarrow})^{-1} \left[\text{P}_{\text{LLL}} \psi_{\text{CB XC}}^{(2)}\right]\\
    = &\int \prod_i^N d\wdown_i d(\wdown_i)^{\ast} \, \text{P}_{\text{LLL}} \psi_{\text{CB XC}}^{(2)} (\zup_1,\dots,\zup_N,\wdown_1,\dots,\wdown_N) \, \psi_{\text{filled LLL}}(\wdown_1,\dots,\wdown_N,\zdown_1,\dots,\zdown_N)\\
    = & \prod_{i<j} (\partial_{\zup_i}-\partial_{\zup_j}) \prod_{i<j} (\zup_i-\zup_j) \, \int \prod_i^N d\wdown_i d(\wdown_i)^{\ast} \, \prod_{i<j}(\partial_{(\wdown_i)^{\ast}}-\partial_{(\wdown_j)^{\ast}}) (\wdown_i-\wdown_j)^{\ast}\\
    & \sum_{\sigma} \sign(\sigma) \prod_i e^{\zup_{\sigma_i} (\wdown_i)^{\ast} / 4\ell^2} \prod_{i<j} (\wdown_i-\wdown_j) \prod_{i<j} (\zdown_i-\zdown_j) \prod_{i,j} (\zdown_i-\wdown_j)\\
    = & \prod_{i<j} (\partial_{\zup_i}-\partial_{\zup_j}) \prod_{i<j} (\zup_i-\zup_j) \prod_{i<j} (\partial_{\zup_i}-\partial_{\zup_j}) \prod_{i<j} (\zup_i-\zup_j)^2 \prod_{i<j} (\zdown_i-\zdown_j) \prod_{i,j} (\zdown_i-2^{-1}\zup_j),
\end{split}
\end{align}
where the additional factor of $2^{-1}$ appears in the interlayer terms because we had to use the modified Bargmann identity
\begin{equation}
    \int dwdw^{\ast} e^{zw^{\ast} / 4\ell^2} f(w) = f(z/2).
\end{equation}

\begin{figure}[h]
    \centering
    \includegraphics[height=0.25\textwidth]{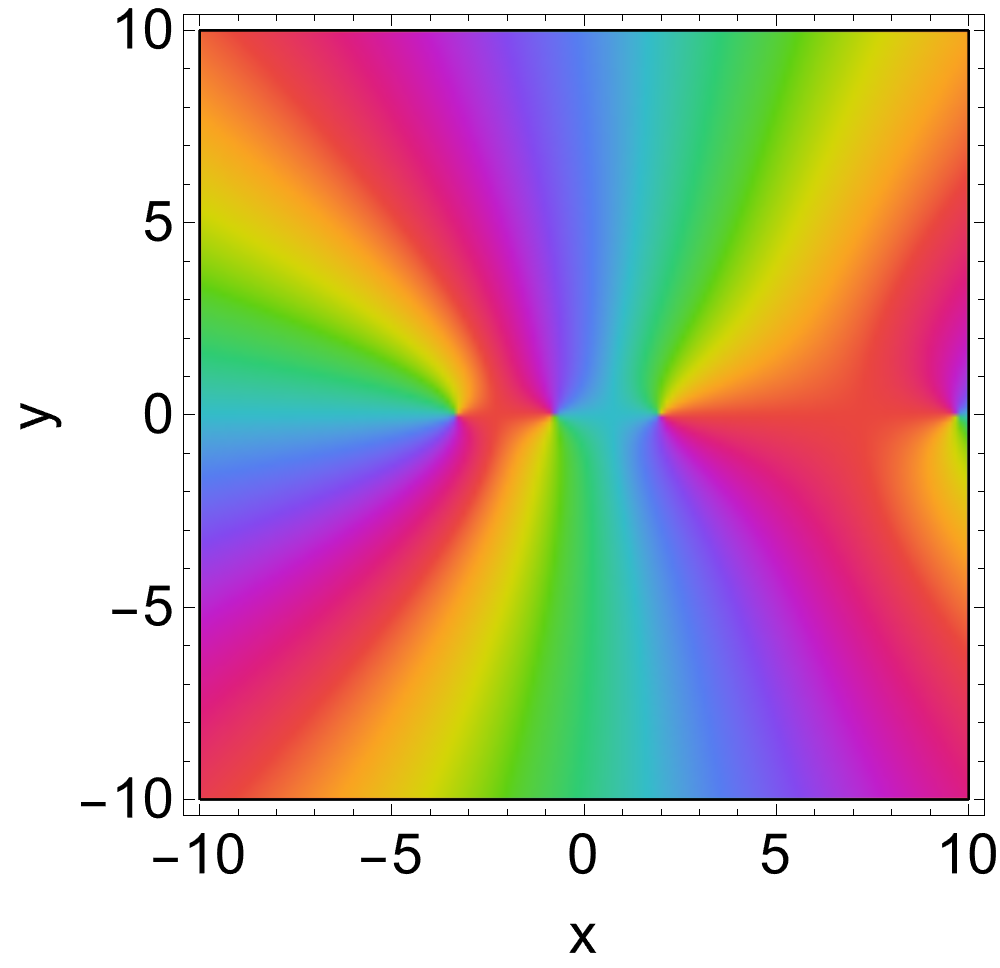}
    \quad
    \includegraphics[height=0.25\textwidth]{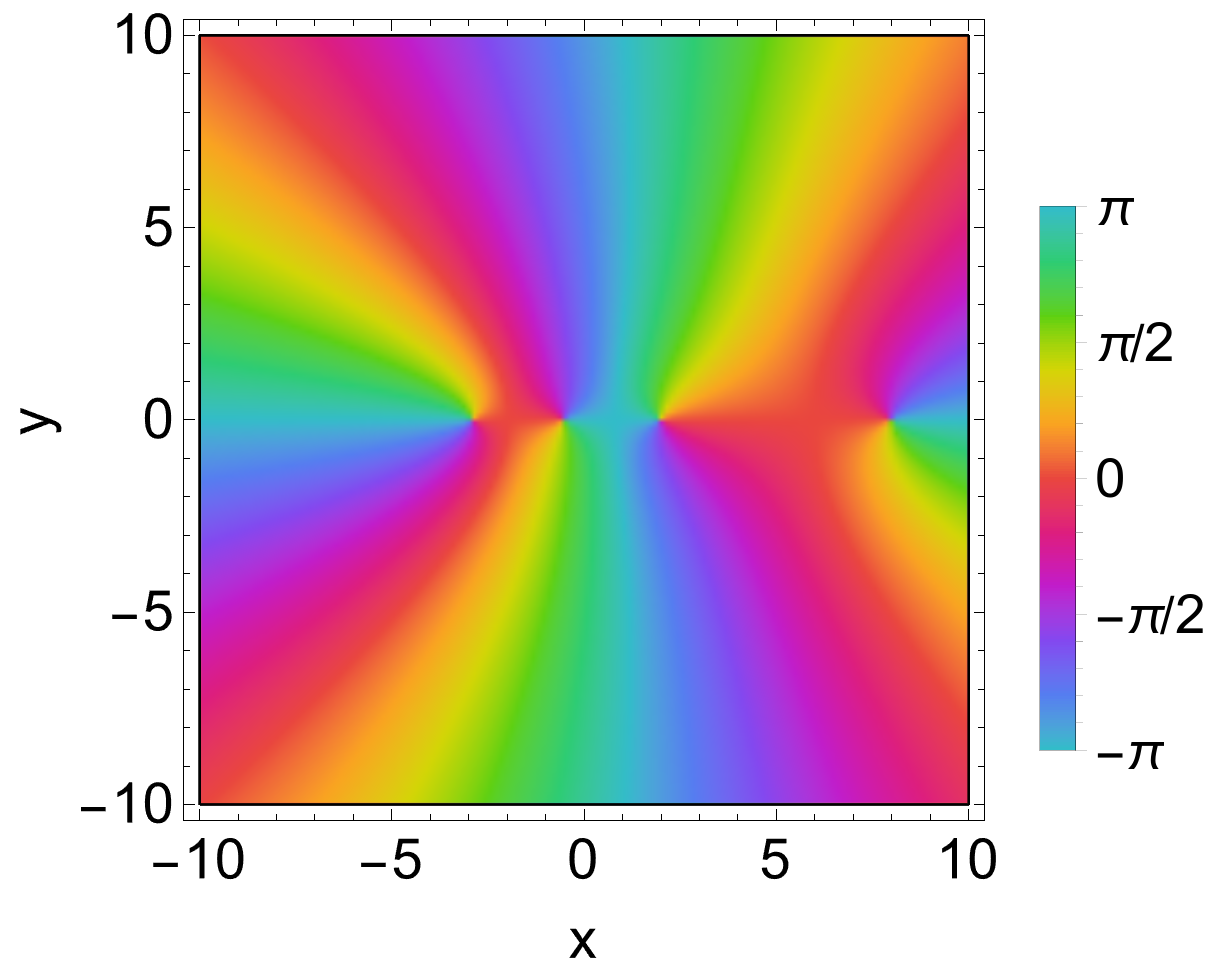}
    \caption{Phases of the wave functions (left) $(\Theta_{\text{PH}}^{\downarrow})^{-1} [\psi_{\text{CB XC}}^{(1)}](\zup_1, \zup_2 = 2, \zdown_1 = -1, \zdown_2 = -2)$ and (right) $(\Theta_{\text{PH}}^{\downarrow})^{-1} [\psi_{\text{CB XC}}^{(2)}](\zup_1, \zup_2 = 2, \zdown_1 = -1, \zdown_2 = -2)$}
    \label{figapp: phase plots phcbxcbiao and phcbxc2}
\end{figure}

In Fig.~\ref{figapp: phase plots phcbxcbiao and phcbxc2}, we compare the phase structures of the PH transformed versions of the CB XC wave functions in Ref.~\cite{Lian2018} and the one proposed here. They again look almost identical. Only the position of the rightmost zero differs a little.

\end{document}